\documentclass{sig-alternate}
\usepackage{color}
\usepackage[dvipsnames]{xcolor}
\usepackage{amsmath}
\usepackage{algorithm}
\usepackage{algpseudocode}
\usepackage{pifont}
\usepackage{url}
\usepackage{hyperref}
\usepackage{setspace}
\usepackage{listings}
\usepackage{fancyvrb}
\usepackage{paralist}
\usepackage{subcaption}

\long\def\comment#1{}

\newcommand{\name}{ZenLDA}

\newcommand{\todo}[1]{\textcolor{red}{#1}}

\usepackage{geometry}
\usepackage{mathptmx} % most powerful
\definecolor{mygreen}{rgb}{0,0.6,0}
\definecolor{mygray}{rgb}{0.55,0.35,0.55}
\definecolor{mymauve}{rgb}{0.58,0,0.82}
\usepackage{multirow}

\begin{document}
\title{ZenLDA: An Efficient and Scalable Topic Model Training System on Distributed Data-Parallel Platform}

\numberofauthors{4} %  in this sample file, there are a *total*
\author{
%
% 1st. author
\alignauthor
Bo Zhao\\
       \affaddr{Microsoft Research}\\
       %\affaddr{1932 Wallamaloo Lane}\\
       %\affaddr{Wallamaloo, New Zealand}\\
       \email{v-zhab@microsoft.com}
% 2nd. author
\alignauthor
Hucheng Zhou\\
       \affaddr{Microsoft Research}\\
       %\affaddr{P.O. Box 1212}\\
       %\affaddr{Dublin, Ohio 43017-6221}\\
       \email{huzho@microsoft.com}
%% 3rd. author
\alignauthor
Guoqiang Li\\
%       %\affaddr{Microsoft Research}\\
%       %\affaddr{P.O. Box 1212}\\
%       %\affaddr{Dublin, Ohio 43017-6221}\\
       \email{witgoing@gmail.com}
%% 4th. author
\alignauthor
Yihua Huang\\
       \affaddr{Nanjing University}\\
%       %\affaddr{P.O. Box 1212}\\
%       %\affaddr{Dublin, Ohio 43017-6221}\\
       \email{yhuang@nju.edu.cn}
}

\maketitle


%A category including the fourth, optional field follows...
%\category{D.2.8}{Software Engineering}{Metrics}[complexity measures, performance measures]


%\keywords{ACM proceedings, \LaTeX, text tagging}sparkLDA

\begin{abstract}
\label{s:abstract}

\comment{
There are increasingly demands on building large-scale distributed system for training machine learning models, such as massive topic models or deep neuron networks.
The system design is quite complicated that it must takes into consideration both machine learning factors and system factors, let alone their interleaving interaction.
Instead of designing the system from scratch, we believe that it can be built on top of existing distributed data-parallel systems such as Apache Spark,
which provides distributed data abstraction and expressive APIs to simplify the parallel programming and simultaneously hides the system complexities such as task scheduling and fault tolerance.
Another advantage is that the entire training pipeline, from training data preparation to model training even model inference, can be programmed in one job and executed in the same framework.
}

This paper presents our recent efforts, {\name}, an efficient and scalable Collapsed Gibbs Sampling (CGS) system for Latent Dirichlet Allocation (LDA) training,
which is thought to be challenging that both data parallelism and model parallelism are required because of the Big sampling data with up to billions of documents and Big model size with up to trillions of parameters.
{\name} combines both algorithm level improvements and system level optimizations.
It first presents a novel CGS algorithm that balances the time complexity, model accuracy and parallelization flexibility.
The input corpus in {\name} is represented as a directed graph and model parameters are annotated as the corresponding vertex attributes.
The distributed training is parallelized by partitioning the graph that in each iteration it first applies CGS step for all partitions in parallel, followed by synchronizing the computed model each other.
In this way, both data parallelism and model parallelism are achieved by converting them to graph parallelism.
We revisited the tradeoff between system efficiency and model accuracy and presented approximations such as unsynchronized model, sparse model initialization and ``converged'' token exclusion.
{\name} is built on GraphX in Spark that provides distributed data abstraction (RDD) and expressive APIs to simplify the programming efforts and simultaneously hides the system complexities.
This enables us to implement other CGS algorithm with a few lines of code change.
To better fit in distributed data-parallel framework and achieve comparable performance with contemporary systems,
we also presented several system level optimizations to push the performance limit.
{\name} was evaluated it against web-scale corpus, and the result indicates that
{\name} can achieve about much better performance than other CGS algorithm we implemented, and simultaneously achieve better model accuracy.
The experiments also demonstrates the effectiveness of presented techniques.

\end{abstract}

\section{Introduction}
\label{s:intro}

% LDA
Topic models provide a way to aggregate vocabulary from a document corpus to form latent ``topics''. In particular,
Latent Dirichlet Allocation (LDA)~\cite{lda} is one of the most popular models~\cite{lda,plsa} that
has rich applications in web mining, from News clustering, hot topics mining, search intent mining even to user interests profiling.
Collapsed Gibbs Sampling (CGS) is the most commonly used algorithm in LDA
that samples the latent variables for a word occurrence (token) by integrating out the Dirichlet priors.
% a distributed solution for LDA training is necessary
However, the training with massive corpus is challenging because of high time and space complexity.
Consider a typical web-scale application with millions of documents and words and with thousands of topics, there are billions of parameters.
No single machine can hold such Big corpus data nor Big model size, which motivates a scalable and efficient way of distributing the computation across multiple machines.

% both algorithm and system
Significant progresses were achieved recently in pushing CGS algorithm into limit,
The time complexity is largely reduced from $O(K)$ (standard LDA), to $O(K_w)$ (SparseLDA~\cite{sparseLDA,F+LDA}), $O(K_d)$ (AliasLDA~\cite{aliasLDA,F+LDA}), and even $O(1)$ in LightLDA~\cite{LightLDA}.
On the other hand, different topic modelling systems were designed but with different parallelization choice.
PLDA~\cite{PLDA}, AD-LDA~\cite{adLDA}, and Peacock~\cite{peacock} were build on MPI~\cite{mpi}/OpenMP~\cite{openmp} primitives~\cite{peacock,japan,PLDA,adLDA}
that provides low-level communication library and APIs to express parallelism.
Yahoo!LDA~\cite{yahooLDA} introduces a parameter server abstraction~\cite{mu,peacock,petuum} and let
each machine put its latest update to server and query the server to retrieve recent updates from other machines.
PLDA+~\cite{plda+} makes use of data placement and pipeline processing to greatly reduce communication time.
LightLDA~\cite{LightLDA} was build on Petuum~\cite{petuum} that is also a parameter server but with stale synchronous parallel (SSP) consistency model~\cite{ssp}.

However, an efficient and scalable solution should combine the innovations from both algorithm side and system side.
Contemporary topic modelling systems either focus on algorithm side~\cite{sparseLDA, F+LDA, aliasLDA, FastLDA} that has different sampling methods,
or focus on system side~\cite{peacock, adLDA, SparkLDA}.
These separation efforts makes it hard to port new algorithms on old systems.
Up to now, there is still no a general system that support all different CGS algorithms, let alone a system that supports different models.
Recently, LightLDA~\cite{LightLDA} is the first trial that integrates both algorithm improvement and system optimization.
However, it conflates the learning algorithm and system logic together, which makes it hard to extend.

Contemporary systems are considered as \texttt{customized approach} that they are almost designed from scratch, programmed in native language and run in a dedicated environment.
They repeatedly address the same system challenges, lose generality due to deep customization and are hardly to debug and extend by couple learning and system together.
In this paper, we consider an alternative (\texttt{generalized approach} that bets on existing distributed data-parallel systems~\cite{mapreduce,dryad,spark} and do not need to consider the system complexities such as scheduling, communication and fault tolerance. Another benefit is that entire learning pipeline, from feature engineering to model training, can be programmed in the same framework,
considering that data-parallel system has already been widely adopted in industry for feature engineering from Big raw data.
Hadoop Mahout~\cite{mahout} and Spark~\cite{spark} MLlib~\cite{mllib} have validated such generalized approach,
such as SparkLDA~\cite{SparkLDA} and the official one in MLlib~\cite{mllibLDA}.
However, they are considered to be performed and scaled poorly that may be 10\~{}100X slower than customized systems.
In this paper, we address the performance concern and try to prove that generalized approach can still achieve comparable or even better efficiency and scalability with customized systems.

% change a sampling algorithm is just several lines of code.
{\name} reflects our latest efforts that builds an efficient and scalable CGS system on distributed data-parallel platform.
{\name} combines both algorithm level improvements and system level optimizations.
It first presents a novel CGS algorithm that balances the time complexity, model accuracy and parallelization flexibility.
The input corpus in {\name} is represented as a directed graph and model parameters are annotated as the corresponding vertex attributes.
The distributed training is parallelized by partitioning the graph that in each iteration it first applies CGS step for all partitions in parallel, followed by synchronizing the computed model each other.
In this way, both data parallelism and model parallelism are achieved by converting them to graph parallelism.
We further revisited the tradeoff between system efficiency and model accuracy and presented approximations such as unsynchronized model, sparse model initialization and ``converged'' token exclusion.
To better fit in distributed data-parallel framework,
we also presented several system level optimizations to push the performance limit.
{\name} is built on GraphX in Spark that provides distributed data abstraction (RDD) and expressive APIs to simplify the programming efforts and simultaneously hides the system complexities.
Such generalization approach enables us to implement other CGS algorithms with only a few lines of code change,
specifically, SparseLDA and LightLDA are implemented as evaluation baseline.
The comparison with them against NYTtimes and one real web-scale dataset with about 3 billions tokens shows
that {\name} can achieve about 2X-6X better performance than LightLDA and about 14X speedup than SparseLDA, and simultaneously achieve better model accuracy.
The effectiveness of presented techniques is also evaluated.
And we also conduct scalability experiment against another bigger Bing web chunk data with about 50 billions tokens and run it a multi-tenancy production environment,
the result indicates {\name} has good scalability and industry-strength quality.

\section{Background and Related Work}
\label{s:back}
This section describes the necessary background, including LDA and the corresponding Collapsed Gibbs Sampling (CGS) training algorithm, as well as the description of
state-of-art distributed data-parallel system, Apache Spark and its library GraphX where {\name} is built.

\subsection{LDA}

% LDA background
In LDA, each of $D$ documents is modeled as a
mixture over $K$ latent topics, each being a multi-nomial distribution over $W$ vocabulary
words. In order to generate a new document $d$, LDA first draw a mixing proportion $\theta_{k|d}$ from
a Dirichlet with parameter $\alpha$. For the $w$th word in the document, a topic assignment $z_{wd}$ is
drawn with topic $k$ chosen with probability $\theta_{k|d}$. Then word $x_{dw}$ is drawn from the $z_{dw}$th topic,
with $x_{dw}$ taking on value $w$ with probability $\phi_{w|k}$, where $\phi_{w|k}$ is drawn from a Dirichlet prior
with parameter $\beta$. Finally, the generative process is below:

\begin{equation}
\theta_{k|d} \sim Dir(\alpha), \phi_{w|k} \sim Dir(\beta), z_{dw} \sim \theta_{k|d}, x_{dw} \sim \phi_{w|z_{dw}}
\label{f:1}
\end{equation}

where $Dir(\alpha)$ represents the Dirichlet distribution.
%Figure~\ref{fig:lda} shows the graphical model representation of the LDA model.

\subsection{Collapsed Gibbs Sampling Algorithm}
% LDA training (Collapsed Gibbs Sampling)
Given the observed words $x = {x_{dw}}$, the task of Bayesian inference for LDA is to compute the
posterior distribution over the latent topic assignments $z = {z_{dw}}$, the mixing proportions $\theta_{k|d}$ and the topics $\phi_{w|k}$.
Approximate inference for LDA can be performed either using variational methods~\cite{lda} or Markov chain Monte Carlo (MCMC) methods~\cite{mcmc}.
In the MCMC context, the usual procedure is to integrate out the mixtures $\theta$ and topics $\phi$ in Formula~\ref{f:1} and just sample the latent variables $z$, which exhibits fast convergence.
This procedure is called Collapsed Gibbs Sampling (CGS), where the conditional
probability of $z_{dw}$ is computed as follows:

\begin{equation}
p(z_{dw}=k|z^{\neg{dw}},x_{dw},\alpha,\beta) \propto \frac{N_{x_{w|k}}^{\neg{dw}} + \beta}{W\beta + N_k^{\neg{dw}}} (N_{k|d}^{\neg{dw}} + \alpha)
\label{f:2}
\end{equation}

where the superscript $\neg{dw}$ means the corresponding topic sampled last time is excluded in the count
values, $N_{k|d}$ denotes the number of tokens in document $d$ assigned to topic $k$, $N_{w|k}$ denotes
the number of tokens with word $w$ assigned to topic $k$, and $N_k = \sum_wn_{w|k}$.
Note that $p_k \geq 0$  is unnormalized.

% Asymmetric LDA
Implementations of topic models typically use symmetric Dirichlet priors with fixed concentration parameters.
However, Wallach, etc. ~\cite{asymLDA} found that an asymmetric Dirichlet prior
over the document-topic distributions ($\alpha$) has substantial advantages over a symmetric prior,
and introduce another hyper-parameter $\alpha'$ to approximate that asymmetric prior as $K\alpha \frac{N_k + \frac{\alpha'}{K}}{\sum_kN_k+\alpha'}$: by that CGS sampling formula can be rewritten as:

\begin{equation}
p(z_{dw}=k|...) \propto \frac{N_{w|k}^{\neg{dw}} + \beta}{W\beta + N_k^{\neg{dw}}} (N_{k|d}^{\neg{dw}} + K\alpha \frac{N_k + \frac{\alpha'}{K}}{\sum_kN_k+\alpha'})
\label{f:3}
\end{equation}

% The serial CGS algorithm
Algorithm~\ref{alg:cgs} describes the standard CGS algorithm~\footnote{We skipped the for-each-occurrence loop between line 4 and line 5.}. Note that the processing order of for-loops in line 3 and line 4 can be interchanged. There are two steps in CGS sampling, first is \texttt{constructing} step that computes the sampling probability of each topic $k$ ($K$-dimensional discrete distribution in total),
followed by \texttt{sampling} step that draws a sample $z$ of topic such that $P_r(z = t) \sim p_t$.
The time complexity is $O(D*W*K)$.
Similarly, the space complexity is also extremely high that the storage of input corpus would be $O(D*W)$,
and with $O(W*K)$ for word-topic matrix, $O(D*K)$ for document-topic matrix.

\begin{algorithm}[t]
\begin{algorithmic}[1]
\Procedure{StandardCGS}{}
\For{each epoch $e$}
    \For{each document $d$}
        \For{each word $w$}
            \For{each topic $k$}
                \State p($k$) = $\frac{N_{w|k}^{\neg{dw}} + \beta}{W\beta + N_k^{\neg{dw}}} (N_{k|d}^{\neg{dw}} + \alpha \frac{N_k + \frac{\alpha'}{K}}{\sum_kN_k+\alpha'})$
            \EndFor
            \State $t$ = TopicSampling(p($k$))
            \State update $N_{t|d}$, $N_{t|k}$ and $N_k$ accordingly
        \EndFor
    \EndFor
\EndFor
\EndProcedure
\end{algorithmic}
\caption{Serial standard CGS algorithm.}
\label{alg:cgs}
\end{algorithm}

\subsection{Spark}

%Spark description, and GraphX
Spark is a fast and general engine for large-scale data processing, which was open-sourced as a top-level Apache project at 2013.
It grew fast with a mature community, and is becoming the de-facto big computing engine that has been widely adopted by industry.
Two types of applications that current computing frameworks handle inefficiently can be benefited from RDDs:
iterative algorithms and interactive data mining tools. In both cases, keeping data in memory can improve performance by an order of magnitude.

\comment{
With the characteristics with \texttt{high performance} that can run programs up to 100x faster than Hadoop MapReduce in memory, or 10x faster on disk,
\texttt{ease of use} that supports application written in Java, Scala, Python and R,
\texttt{generality} that combines SQL and DataFrames, Spark streaming, GraphX for graph computing and MLlib for machine learning,
and \texttt{runs everywhere} that includes Hadoop, Mesos, standalone or the cloud,
Spark grew fast with a mature community, and is becoming the de-facto big computing engine that has been widely adopted by industry.
}

\textbf{RDD abstraction.}
% technique of Spark
Spark improves the distributed data-parallel systems such as MapReduce, Hadoop, Dryad by providing a Resilient Distributed Datasets (RDDs) abstraction,
which is an efficient, general-purpose and fault-tolerant abstraction for sharing data in cluster applications.
Essentially, RDD represents an immutable, partitioned collection of elements that can be operated on in parallel.
The RDD element can be any type, from primitive types to complex classes.
RDDs is immutable, and is offered with APIs that support coarse-grained transformations that transform RDDs and actions that return result.
lets them recover data efficiently using lineage that tracks how to re-compute lost data from previous RDDs.
Users can explicitly cache an RDD in memory or disk across machines
and reuse it in successive computing.
In addition, Spark supports two restricted types of shared variables, accumulator that workers can only ``add'' to using an associative operation and only the driver can read,
and broadcast variable that create a ``broadcast variable'' object that wraps the value and ensures that it is only copied to each worker once.
Spark core itself is written in Scala language, and each RDD is represented by a Scala object.
High-level languages such as Java, Python and R are also supported in SparkR as a light-weight frontend.
Therefore, developer can easily write learning applications like single-box environments.

To use Spark, developers write a driver program that implements the high-level control flow of their application
and launches various RDD operations in parallel. These operations are invoked by passing a function to apply on
a RDD. Driver is responsible for scheduling the tasks and coordinating the worker execution.

\textbf{Machine learning.}
% ML related components
It is unsurprised that there are already progresses on learning in Spark.
MLlib is Spark¡¯s machine learning (ML) library. Its goal is to simplify practical machine learning programming, by provides high-level representations \texttt{Vector/Matrix/DataFrames} on top of RDDs.
MLlib consists of common learning algorithms and utilities, including classification, regression, clustering, collaborative filtering, dimensionality reduction,
as well as lower-level optimization primitives and higher-level pipeline APIs.
It is noteworthy that there already are two LDA implementations, including expectation-maximization (EM)~\cite{mllibLDA} on the likelihood function and variational inference based online training.

Besides data/model parallelism, graph parallelism is also inherent to learning algorithm and is widely used to parallelize the training process.
GraphX extends the Spark RDD by introducing a new \texttt{Graph} abstraction: a directed graph with properties attached to each vertex and edge.
To support graph computation, GraphX exposes a set of fundamental operators (e.g., subgraph, joinVertices, and aggregateMessages) as well as an optimized variant of the Pregel API.
In addition, GraphX includes a growing collection of graph algorithms and builders to simplify graph analytics tasks.

\section{{\name} Design}
\label{s:design}
In this section, we first describe the serial CGS algorithm in {\name} that has different formula decomposition,
then followed by the comparison with existing approaches.

\subsection{Sampling approach.}
\label{sec:cgs}
Another dimension in design space is the choice of decomposition of Formula~\ref{f:3}, which costs large proportion of execution time in one iteration.
Different formula decompositions have different sampling characteristics.
There are three major considerations in {\name}: 1). whether the decomposed part is loop invariant or with negligible change?
For example, $\frac{\alpha_k*\beta}{N_k+W\beta}$ is loop invariant while $N_{k|d} * N_{w|k}$ changes significantly.
2). whether the decomposed part is sparse with respect to topic $k$? Sparse part has less computing complexity as well as less memory consumption.
For example, $N_{w|k}*\alpha$ is sparse since $N_{w|k}$ is sparse, and the computing complexity is $O(K_w)$.
3). whether or not the approximation is permitted in computing topic probability that does not compromise sampling accuracy?
It is reasonable to the approximation on formula parts with less value proportion would have less deviation errors in total.
For instance, $N_{k|d} * N_{w|k}$ has largest value, while $\alpha_k * \beta$ is the smallest.
It is thus unnecessary to compute the less important part every time, including $N_{k|d}*\beta$ and $N_{w|k}*\alpha$.

\noindent\textbf{{\name} decomposition.}
{\name} chooses a different decomposition with $\frac{\alpha_k*\beta}{N_k+W\beta} + \frac{N_{wk}*\alpha_k}{N_k+W\beta} + \frac{N_{kd}*(N_{wk}+\beta)}{N_k+W\beta}$, which has the following benefits compared with other approaches:
\begin{compactitem}
  \item $\frac{\alpha_k*\beta}{N_k+W\beta}$ is only computed once and reused afterwards in an iteration.
        And an alias table~\cite{aliasLDA,F+LDA}, $gTable$, is created accordingly, thus $O(1)$ sampling complexity is achieved.
        Approximation happens here since $N_k$ changes for each sampling, that is why SparseLDA adopts linear search based sampler which has $O(K)$ sampling complexity.
  \item $\frac{N_{wk}*\alpha_k}{N_k+W\beta}$ is also approximated that it is pre-computed and reused for the same word ($w$).
        Similarly, the alias table ($wSparse$) is created accordingly.
        The lifecycle of this alias table in {\name} is reduced with word-by-word process order, that all tokens of the same word are grouped and sampled together.
        Recall that the corresponding topic ($k$) sampled last time should be excluded in the count values ($N_{w|k}$) for current sampling, i.e, $N_{w|k}$ should be subtracted by one for that $k$.
        However, such subtraction is skipped since there is no information on which topic should be subtracted during the pre-computing.
        We apply remedy by resampling with a probability of $\frac{1}{N_{w|k}}$ if the sampled topic is equal to the topic sampled last time. This is especially useful when $N_{w|k}$ is small that is close to 1.
  \item Only $\frac{N_{kd}*(N_{wk}+\beta)}{N_k+W\beta}$ is computed for each token with $O(K_d)$ time complexity.
        And a cumulative distribution function (CDF) is created and the corresponding sampling complexity is $O(logK_d)$.
        It is noteworthy that it is only computed once for different occurrences of the same word in the same document.
        Thereby similarly, this did not subtract 1 for $N_{k|d}$ and $N_{w|k}$, thus the $1*\beta$ should be excluded from $N_{k|d}*\beta$ and $N_{k|d} + N_{w|k} - 1$ should be excluded from $N_{k|d}*N_{w|k}$.
        Therefore, resampling is applied for remedy with probability of $\frac{1}{N_{k|d}} + \frac{N_{k|d} + N_{w|k} - 1}{N_{k|d}*N_{w|k}}$ if the sampled topic is equal to the topic sampled last time
        ~\footnote{We actually compute this decomposed part with 1 is subtracted if only one occurrence for the same (word, document) pair.}.
\end{compactitem}

\noindent\textbf{{\name}Hybrid decomposition.}
$\frac{\alpha*\beta}{N_k+W\beta} + \frac{N_{wk}*\alpha}{N_k+W\beta} + \frac{N_{kd}*(N_{wk}+\beta)}{N_k+W\beta}$ is better than
$\frac{\alpha*\beta}{N_k+W\beta} + \frac{N_{kd}*\beta}{N_k+W\beta} + N_{wk}(\frac{N_{kd}+\alpha}{N_k+W\beta})$ (used in SparseLDA~\cite{sparseLDA,peacock}),
since decomposed part $N_{wk}(\frac{N_{kd}+\alpha}{N_k+W\beta})$ in latter one has complexity of $O(K_w)$, which is worse than $O(K_d)$ in former one,
consider that word-topic array is generally more dense than document-topic array.
However, the long-tail words may have less occurrences than the document length, thus the corresponding word-topic array may be more sparse that $K_w < K_d$.
We further provide a hybrid sampling approach, \texttt{{\name}Hybrid}, that alternates the formula decomposition between them,
that we choose the former one for tokens with more sparse document-topic array;
otherwise, we choose the latter one for tokens with more sparse document-topic array.
Note that $\frac{N_{kd}*\beta}{N_k+W\beta}$ would have significant change, we adopt F+ tree based sampler that has $O(logK_d)$ complexity.
To minimize the lifecycle of alias table that corresponds to the second decomposed part, the tokens should be grouped according to vertex that has larger degree, and tokens in a group are processed together.

\subsection{Algorithm}
\noindent\textbf{{\name} algorithm.}
The specific serial algorithm of CGS training in {\name} is described in Algorithm~\ref{alg:cgs-zen}.
We skipped the algorithm for {\name}Hrbrid that is with a natural extension.
Compared with standard CGS that has $O(K)$ complexity, {\name} significantly reduces the complexity into $O(min(K_d, K_w))$.

% parallelization  (data parallelism and model parallelism)
There are multiple factors that constitute large design space when consider to parallelize the CGS across multiple machines.

\comment{
\begin{figure}[t]
\vspace*{-1ex}
\begin{center}
\begin{lstlisting}[mathescape=true]
CGSTraining() {
  for (each epoc $e$) {
    for (each topic $k \in K$)
      $gDense \gets \frac{\alpha_k*\beta}{N_k+W\beta}$
    $gTable \gets createAliasTable(gDense)$
    for (each word $w \in W$) {
      for (each topic $k \in K_w$)
        $wSpase \gets \frac{N_{w|k}*\alpha_k}{N_k+W\beta}$
      $wTable \gets createAliasTable(wSparse)$
      for (each edge $e = (w, d_i) \in E, d_i \in D$) {
        for (each topic $k \in K_d$)
          $dSparse \gets \frac{N_{k|d}*(N_{w|k}+\beta)}{N_k+W\beta}$
        $dTable \gets createAliasTable(dSparse)$
        for (each token $t \in {e_{dw}}$) {
          $z_t \gets sample(gTable, wTable, dSparse)$
          Update $N_{k|d}, N_{w|k}, N_k$
        }
      }
    }
}}
\end{lstlisting}
\vspace{-0.4cm}
\caption{The single-box CGS training algorithm in {\name}.}
\label{alg:cgs-zen}
\end{center}
\vspace*{-3ex}
\end{figure}
}

\begin{algorithm}[t]
\begin{algorithmic}[1]
    \State{\textbf{Input:} The set of edges $E$; the set of words $W$ and documents $D$ in a partition $P$.}
    \State{\textbf{Output:} Sample new topic for each edge and update the model state.}
    \Procedure{CGSTraining}{}
        \For{each epoc $e$}
        \For{each topic $k \in K$}
            \State {$gDense \gets \frac{\alpha_k*\beta}{N_k+W\beta}$}
        \EndFor
        \State{$gTable \gets createAliasTable(gDense)$}
        \For{each word $w \in W$}
            \For{each topic $k \in K_w$}
                \State{$wSpase \gets \frac{N_{w|k}*\alpha_k}{N_k+W\beta}$ }
            \EndFor
            \State{$wTable \gets createAliasTable(wSparse)$}
            \For{each edge $e = (w, d_i) \in E, d_i \in D$}
                \For{each topic $k \in K_d$}
                    \State {$dSparse \gets \frac{N_{k|d}*(N_{w|k}+\beta)}{N_k+W\beta}$}
                \EndFor
                \State{$dTable \gets createAliasTable(dSparse)$}
                \For{each token $t \in {e_{dw}}$}
                    \State{$z_t \gets sample(gTable, wTable, dSparse)$}
                    \State{Update $N_{k|d}, N_{w|k}$ and $N_k$}
                \EndFor
            \EndFor
        \EndFor
        \EndFor
    \EndProcedure
\end{algorithmic}
\caption{The single-box CGS training algorithm in {\name}.}
\label{alg:cgs-zen}
\end{algorithm}

\subsection{Related work in CGS algorithm}
\label{s:relcgs}
\begin{table*}
\centering
\renewcommand\arraystretch{1.3}
\begin{tabular}{|l|c|c|c|} \hline
            & {\name} & {\name}Hybrid & AliasLDA \\
\hline
\scriptsize{Decomposition}
              & $\frac{\alpha_k*\beta}{N_k+W\beta} + \frac{N_{wk}*\alpha_k}{N_k+W\beta} + \frac{N_{kd}*(N_{wk}+\beta)}{N_k+W\beta}$
              & $\frac{\alpha_k*\beta}{N_k+W\beta} + \frac{N_{kd}*\beta}{N_k+W\beta} + N_{wk}(\frac{N_{kd}+\alpha_k}{N_k+W\beta})$
              & $\alpha_k(\frac{N_{wk}+\beta}{N_k+W\beta}) + N_{kd}(\frac{N_{wk}+\beta}{N_k+W\beta})$ \\
\scriptsize{Sampler} & Alias\hspace{21pt} Alias\hspace{26pt} CDF & Alias\hspace{21pt} Alias\hspace{26pt} CDF & Alias\qquad Alias \\
\scriptsize{Fresh} & no\hspace{30pt} no\hspace{38pt} yes &no\hspace{30pt} no\hspace{38pt} yes & no \hspace{26pt} yes \\
\scriptsize{Computing} & $O(1)$\hspace{20pt} $O(1)$\hspace{21pt} $O(K_d)$ & $O(1)$\hspace{20pt} $O(1)$\hspace{21pt} $O(min(K_d,K_w))$  & $O(1)$\qquad $O(K_d)$ \\
\scriptsize{Sampling} & $O(1)$\hspace{20pt} $O(1)$\hspace{21pt} $O(logK_d)$ & $O(1)$\hspace{20pt} $O(1)$\hspace{21pt} $O(min(logK_d,logK_w))$  &  $O(\#MH)$\qquad $O(\#MH)$ \\
\hline
\scriptsize{Process Order} & Word-by-Word & Doc-by-Doc & Doc-by-Doc \\
\scriptsize{Approximation} & yes & yes & no \\
\hline
\end{tabular}

\begin{tabular}{|l|c|c|c|c|} \hline
            & LightLDA & F+LDA & F+LDA & SparseLDA \\
\hline
\scriptsize{Decomposition}
              & $\frac{N_{wk}+\beta}{N_k+W\beta}\enspace * \enspace N_{kd}+\alpha$
              & $\alpha(\frac{N_{wk}+\beta}{N_k+W\beta}) + N_{kd}(\frac{N_{wk}+\beta}{N_k+W\beta})$
              & $\beta(\frac{N_{kd}+\alpha}{N_k+W\beta}) + N_{wk}(\frac{N_{kd}+\alpha}{N_k+W\beta})$
              & $\frac{\alpha*\beta}{N_k+W\beta} + \frac{N_{kd}*\beta}{N_k+W\beta} + N_{wk}(\frac{N_{kd}+\alpha}{N_k+W\beta})$ \\
\scriptsize{Sampler} & Alias\qquad Alias & F+tree\qquad BSearch & F+Tree\qquad BSearch & LSearch\qquad LSearch\qquad LSearch  \\
\scriptsize{Fresh} & no \hspace{26pt} yes & yes\hspace{26pt} yes & yes\hspace{26pt} yes & yes \hspace{26pt} yes\hspace{26pt} yes \\
\scriptsize{Computing} & $O(1)$\hspace{26pt} $O(1)$ & $O(logK)\qquad O(K_d)$ & $O(logK)\qquad O(K_w)$ &$O(1)$\qquad $O(1)$\qquad $O(K_w)$\\
\scriptsize{Sampling} & $O(\#MH)$\qquad $O(\#MH)$  & $O(logK)\qquad O(logK_d)$ & $O(logK)\qquad O(logK_w)$  & $O(K)$\qquad $O(K_d)$\qquad $O(K_w)$ \\
\hline
\scriptsize{Process Order} & Word-by-Word & Word-by-Word & Doc-by-Doc & Doc-by-Doc  \\
\scriptsize{Approximation} & no & no & no & yes \\
\hline
\end{tabular}
\caption{Comparison of different LDA sampling approaches. Note that two decompositions are alternately used in {\name}Hybrid, another one is listed in {\name}.}
\label{t:lda-comp1}
\end{table*}

Table~\ref{t:lda-comp1} depicts the detailed summary on comparison among different CGS approaches. Besides the difference in decomposition,
this table also list the difference on which \texttt{sampler} is used, whether it is \texttt{fresh} that the formula is computed for each token, whether \texttt{approximation} is applied,
the corresponding \texttt{computing complexity} if computing is needed, the \texttt{sampling complexity}, and the \texttt{process order} applied in CGS step.

SparseLDA~\cite{sparseLDA} is the first sampling method which considered decomposing $p_k$ into a sum of sparse vectors and a dense vector.
In particular, it considers a three-term decomposition as:
\begin{center}
$\frac{\alpha*\beta}{N_k+W\beta} + \frac{N_{kd}*\beta}{N_k+W\beta} + N_{wk}(\frac{N_{kd}+\alpha}{N_k+W\beta})$,
\end{center}
where the first term is dense, the second term is sparse
with $K_d$ non-zeros, and the third term is sparse with $K_w$ non-zeros.
As SparseLDA follows the document-by-document sequence, very few elements will be changed for the first two terms at each CGS step.
Linear search~\cite{F+LDA} (LSearch) is applied to all of these three terms in both SparseLDA implementations (Yahoo!LDA~\cite{yahooLDA} and Mallet LDA~\cite{sparseLDA}),
which makes the sampling complexity with $O(K)$, $O(K_d)$ and $O(K_w)$ for these three terms, respectively.

AliasLDA~\cite{aliasLDA} considers the following decomposition of p:
\begin{center}
$\alpha(\frac{N_{wk}+\beta}{N_k+W\beta}) + N_{kd}(\frac{N_{wk}+\beta}{N_k+W\beta})$,
\end{center}
Instead of the exact multinomial sampling, AliasLDA considers a proposal distribution $q_k$ with a very efficient generation routine
and performs a series of Metropolis-Hasting (MH) steps using this proposal to simulate the true distribution $p_k$.
In particular, the proposal distribution is constructed using the latest second term and a stale version of the first term.
For both terms, Alias method is applied. $\#MH$ steps decides the quality of
the sampling results. The overall amortized cost for each CGS step is $O(K_d + \#MH)$. Note the initialization cost
$O(K)$ for the first term can be amortized. Therefore, AliasLDA reduces the amortized cost of each step to $O(K_d)$.

LightLDA~\cite{LightLDA} is a recently proposed approach that develops an Metropolis Hastings sampler with different constructed proposals that
combines the document-proposal ($N_{k|d} + \alpha$) and word-proposal ($\frac{N_{wk}+\beta}{N_k+W\beta}$) into a ``cycle proposal''.
The word-proposal is pre-computed and Alias method is applied, thus the $O(K)$ complexity is amortized to be $O(1)$.
And $O(1)$ complexity is also achieved by a lookup table with document length that stores the corresponding topic for its word occurrences.

F+LDA~\cite{F+LDA} has two variants that with different decompositions. The document-by-document process order has the decomposition as:
\begin{center}
$\beta(\frac{N_{k|d}+\alpha}{N_k+W\beta}) + N_{w|k}(\frac{N_{k|d}+\alpha}{N_k+W\beta})$,
\end{center}
For each document, a sample step only changes two topic counter for $N_{k|d}$ in the first term, so they adopts F+tree sampling for the first term that with $O(logK)$ complexity for both updating and sampling.
Correspondingly, different word in the same document has different $N_{w|k}$ for the second term, and the complexity is $O(K_w)$.
Similarly, the word-by-word process order has opposite decomposition and properties:
\begin{center}
$\alpha(\frac{N_{w|k}+\beta}{N_k+W\beta}) + N_{k|d}(\frac{N_{w|k}+\beta}{N_k+W\beta})$,
\end{center}

It is worth to compare with LightLDA, since the complexity of {\name} is better than other alternatives but could be worse than LightLDA~\cite{LightLDA}.
First, LightLDA needs an extra lookup table that stores the correspondence of a token and its sampled topic (analogous to edge and its attribute in {\name}).
Instead of directly read $N_{k|d}$, LightLDA samples the lookup table to simulate $N_{k|d}$, thus the complexity is reduced from $O(K_d)$ to $O(1)$.
However, tt requires data is partitioned in a document-wise way, otherwise the sample result would be inaccurate.
This limits the exploration of better partition approaches (Section~\ref{s:para}).
Second, a MH-step will compute the true probability (Formula~\ref{f:3}) of the sampled topic, $O(1)$ complexity can only be achieved when dense vector or hash table is used,
this will result in high memory consumption.
Otherwise, MH-step could result in $O(max(K_w, K_d))$ complexity to get value from $N_{w|k}$ and $N_{k|d}$ if the sparse data structure is used.

\section{{\name} Parallelization}
\label{s:design}
In this section, we first discuss the parallelization design in {\name}, followed by other useful utilities {\name} provided.

\subsection{Parallelization design in {\name}}
\label{s:para}
Consider a typical web-scale application with 100 millions of documents and words, and with a large number of topics (typically on the order of thousands), where there are almost trillions of parameters. No single machine can hold the entire Big corpus data nor the Big model size.
This made single machine solution impossible, which motivates a scalable and efficient way of distributing the computation across multiple machines.
However, the design is challenging that a typical web-scale LDA training requires both data parallelism and model parallelism and involves hundreds of machines.
In this section, we will discuss multiple design dimensions to parallelize the {\name} across multiple machines.

\noindent\textbf{Graph based data and model representation.}
In stead of represented the data (input corpus) and model (word-topic and document-topic) as matrix~\footnote{Note that both data and model are sparse.},
% graph model
{\name} represents data as a directed bipartite graph that is the dual representation of sparse matrix.
Figure~\ref{fig:lda-graph} depicts the graph representation of a corpus with three words ($w_1,w_2,w_3$) and documents ($d_1,d_2,d_3$).
The graph has two kinds of vertices, word vertices and document vertices.
An edge exists from word vertex to document vertex only if that word is occurred in the document.
This corpus in LDA is a natural graph~\cite{powergraph} like many other natural language processing problems,  where the graph have highly skewed power-law degree distributions.
% implementation in Spark
This graph representation can be naturally mapped to \texttt{Graph} in GraphX.
Note that the corpus graph is treated as directed graph just because {\texttt{Graph} in GraphX is directed, the direction is actually meaningless.
The edges in GraphX is grouped in a partition according to the source vertex ID (word vertex).

The model parameter word-topic matrix is split in a word-wise fashion that each word vertex is attached with the corresponding word-topic array ($N_{w|k}$) as attributes.
Similarly, the document-topic matrix is also split in a document-wise fashion and each document vertex is attached with the corresponding document-topic array ($N_{k|d}$).
Both word-topic and document-topic array are sparse that not all topics are sampled,
and they are becoming more and more sparse as the training converged.
Relatively speaking, a long-tail word may have more sparse word-topic array than a hot word;
and document-topic array maybe more sparse than word-topic array since a word may have more occurred tokens than the average document length.
The current topic ($Z_{dw}$) of a word occurrence (token) $w_{dw}$ is annotated as the corresponding edge attribute.
It is noteworthy that the edge attribute is an array since there may be multiple occurrences of the same word in one document.
And the global state $N_k = \sum_dN_{k|d} = \sum_wN_{w|k}$  records the total number of tokens (edges) been sampled as topic $k$, which
is computed by aggregating the $N_{k|d}$ from all document vertices (or $N_{w|k}$ from all word vertices).

\begin{figure}
\centering
\includegraphics[width=0.4\textwidth]{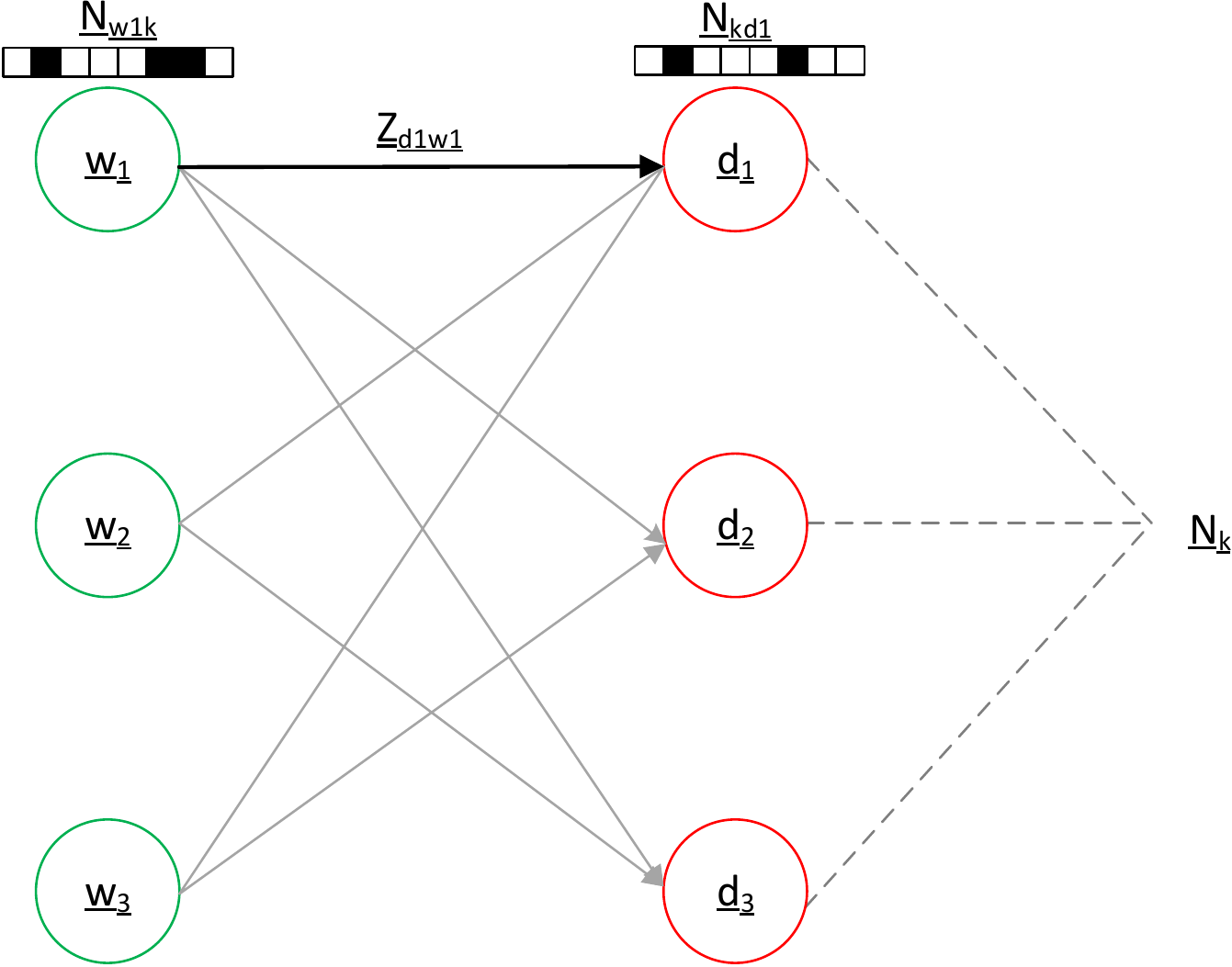}
\caption{Graph based CGS abstraction.}
\label{fig:lda-graph}
\end{figure}

\noindent\textbf{Partition approach.}
%% parallelization
The distributed parallelism is achieved by partitioning the graph into multiple partitions (described in next paragraph), and workers apply CGS process in Algorithm~\ref{alg:cgs-zen} for all partitions in parallel,
followed by synchronizing the model state at the end of iteration. In this way, data parallelism and model parallelism are achieved~\cite{peacock}, where the model is also partitioned and distributed across workers.

Partition strategy that determines how to partition the corpus and model plays crucial impact on system performance.
The improper partition would result in load imbalance and large network communication.
Compared with (sparse-)matrix based representation that can only be partitioned in a ``rectangle'' way, graph has more freedom for partitioning choice.
There are two partition strategies, \texttt{edge-cut} that tries to evenly assign the vertices to machines by cutting the edges.
\texttt{vertex-cut}~\cite{powergraph} that tries to evenly assign the edges to machines by cutting the vertices.
PowerGraph~\cite{powergraph} pointed out that vertex-cut can achieve better performance than edge-cut, especially for power-law graphs.
And the workload of a machine in vertex-cut is determined by the number of edges located in that machine, and the total communication cost is proportional to the number of mirrors of the vertices.
However, the power-law distributions in corpus graph makes the partitioning challenging~\cite{dbh}.
GraphX currently only supports vertex-cut method, and provides three partitioning approaches:
\texttt{RandomVertexCut} that assigns edges to partitions by hashing the source and destination vertex IDs;
\texttt{EdgePartition1D} that assigns edges to partitions using only the source/destionation vertex ID, collocating edges with the same source/destionation;
and \texttt{EdgePartition2D} that assigns edges to partitions using a 2D (``rectangle'') partitioning of the sparse edge adjacency matrix, guranteeing a $2*\sqrt{numParts}$ bound on the number of vertex replication.

Xie, .etc~\cite{dbh} presented degree-based hashing (DBH) partition method
that can achieve lower communication cost than existing methods and can simultaneously guarantee good workload balance.
The theoretical bounds on the communication cost and workload balance can also be derived.
DBH is also vertex-cut that it first applies randomized hash function to evenly assign vertices to partitions,
then assigns an edge ($(v_i, v_j)$) to partition that contains its source or destination whose degree is less.
In other words, the vertex with larger degree is cut that it would have multiple replicas.
DBH shares the same insights with PowerLyra~\cite{PowerLyra} that locality matters for low-degree vertex thus it places all edges related to this vertex together,
while parallelism matters for high-degree vertex thus it favors to cut high-degree vertex.
However, DBH only considers the relative size between source degree and destination degree,
without considering their absolute value.
Consider the case where both source and destination degree are small (smaller than a threshold value), it is not reasonable to still correspond the edge to vertex with lower degree, but should be the vertex with higher degree.
In {\name}, we improved DBH as DBH+, and the algorithm is listed in Algorithm~\ref{alg:dbh+}.

\begin{algorithm}[t]
\begin{algorithmic}[1]
    \State{\textbf{Input:} The set of edges $E$; the set of vertices $V$; the number of machines $p$.}
    \State{\textbf{Output:} The assignment $P(e) \in [p] partitions$ for each edge $e$.}
    \Procedure{DBHPlus}{}
        \For{each $v \in V$}
            \State {$P(v) = hash(v)$}
        \EndFor
        \State {count the degree $d_i$ for each $i \in V$ in parallel}
        \For{each $e = (v_i,v_j) \in E$}
            \If {$max(d_i,d_j) < threshold$}
                \If {$d_i \leq d_j$}
                    \State {$P(e) = P(d_j)$}
                \EndIf
            \Else
                \If {$d_i \leq d_j$}
                    \State {$P(e) = P(d_i)$}
                \Else
                    \State {$P(e) = P(d_j)$}
                \EndIf
            \EndIf
        \EndFor
    \EndProcedure
\end{algorithmic}
\caption{DBH: an improved Degree-based hashing (DBH) algorithm.}
\label{alg:dbh+}
\end{algorithm}

We actually explored many other partitioning strategies, such as different greedy algorithms~\cite{PowerLyra} and iterative algorithm~\cite{website:Giraph}.

\noindent\textbf{Synchronization approach.}
After partitioning and distributing the data/model, the remaining thing to consider is the synchronization among machines.
In theory, the update of topic assignment $Z_{dw}$ can not be performed concurrently with the update
of any other topic assignment $Z_{d'w'}$, with conflicts on $N_k$ and possible conflicts on $N_{w|k}$ or $N_{k|d}$
The conflicts must be guaranteed using locks~\footnote{Topic level parallelism exists that the topic probability computing for each topic ($p(k)$) can be parallelized without any locks.}, which is costly and is hardly implemented in distributed environment.
The good news is that this dependence is weak, given the typically large number of word tokens compared to the number of machines.
If two processors are concurrently sampling, but sampling different words in different documents
(i.e., $w_{dw} \neq w{d'w'}$, then concurrent sampling will be very close to sequential sampling that the only affected state is $N_k$.
This means that we can still achieve convergence by relaxing these locks.
There are three possible design choices, either only $N_{k|d}$ or $N_{w|k}$ is strictly synchronized, or none of them are not synchronized.
This is more suitable for distributed data-parallel processing that communication happens only across the stage boundary.
\comment{ % related work
AD-LDA~\cite{adLDA}, SparseLDA~\cite{sparseLDA}, AliasLDA~\cite{aliasLDA} and SparkLDA~\cite{sparkLDA} permit the asynchronized update on $N_{w|k}$ and $N_k$.
{\name} supports all these three synchronization approaches.
Unlike SparkLDA~\ref{sparkLDA} and LightLDA~\ref{lightLDA} that further block a partition into the mini-batches in a ``conjugated'' way~\footnote{If the partition is document-wise, then the mini-batch is word-wise, or vice versa.}, and synchronization happens across mini-batches to reduce the state staleness.
}
The first two can be achieved by choosing \texttt{EdgePartition1D} method that either all edges corresponding to a word or a document are located in the same partition,
where $N_{k|d}$ or $N_{w|k}$ is not synchronized.
However, \texttt{EdgePartition1D} would result in data severely skew, that even the number of documents is even distributed but the number of words is not, or vice versa.
Therefore, {\name} further aggressively relaxes the dependency that both model states are updated independently,
this asynchronization approach enables us to choose any possible partition methods that has better load balance with less network communication, such as DBHPlus presented above.
Furthermore, even inside a partition, the CGS process result is only used to update the edge attribute ($Z_{dw}$), and only update the vertex attribute ($N_{k|d}$, $N_{w|k}$) at the end,
i.e., line 21 in Algorithm~\ref{alg:cgs-zen} is moved to at the epoch. This will largely reduce the lock cost if multi-threaded is enabled inside a partition.
The side effect is that $N_{k|d}$, $N_{w|k}$, and $N_k$ are stale with value computed in last iteration.
\comment{\todo{why they would not affect convergence?}}

\begin{figure}
\centering
\includegraphics[width=0.5\textwidth]{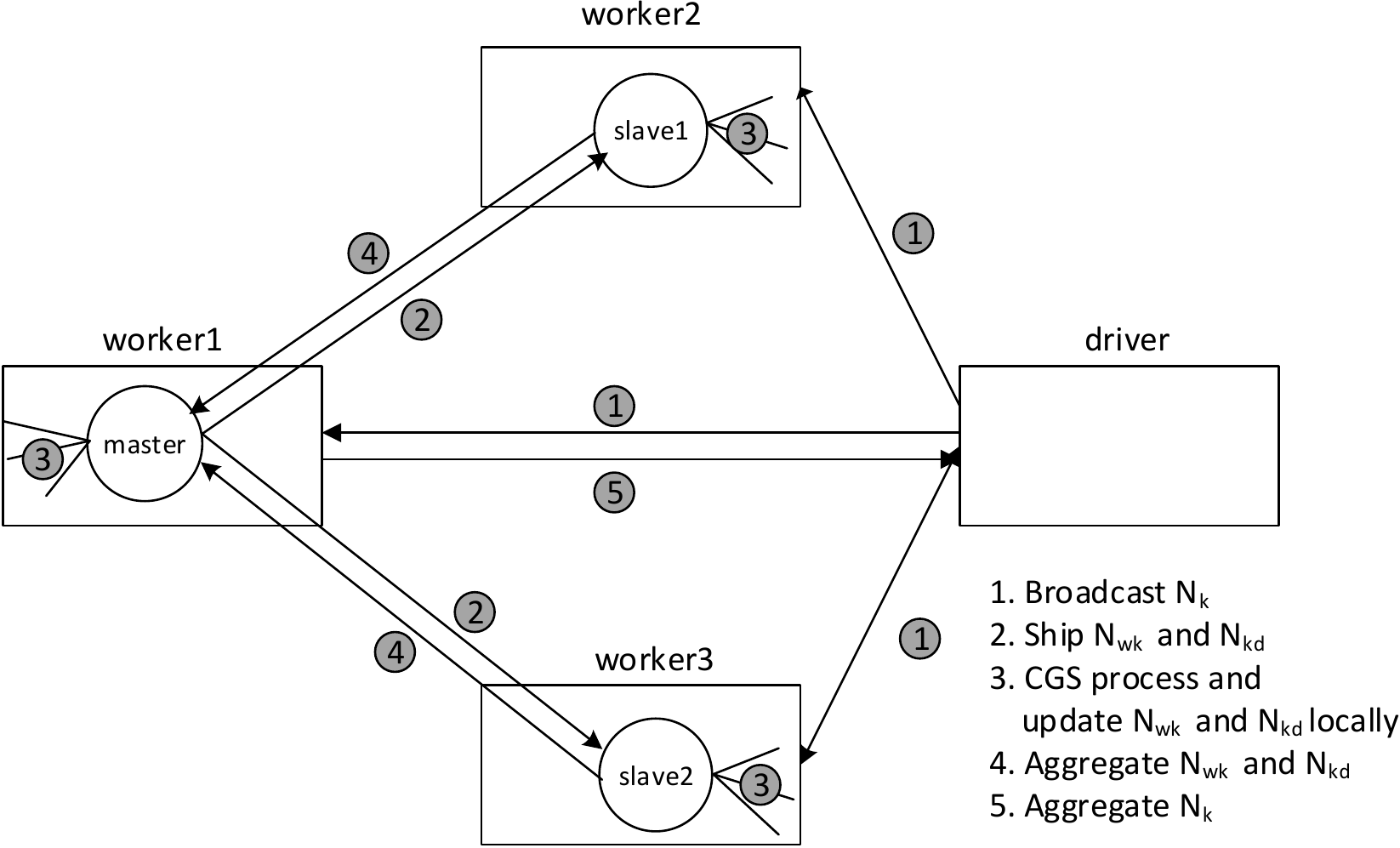}
\caption{{\name} workflow in an iteration.}
\label{fig:workflow}
\end{figure}
\noindent\textbf{{\name} workflow.}
In conclusion, the CGS workflow of one iteration in {\name} is illustrated in Figure~\ref{fig:workflow}. There are five steps:
1). driver broadcasts $N_k$ to all workers. 2). the vertex master ships  the model state ($N_{k|d}$ or $N_{w|k}$) to all of the corresponding vertex slaves.
3). workers apply Algorithm~\ref{alg:cgs-zen} in parallel and update the $N_{k|d}$ or $N_{w|k}$ locally at the end.
4). at the end of an iteration, vertex master aggregates $N_{k|d}$ or $N_{w|k}$ all local updates from workers.
5). driver aggregates $N_k = \sum_wN_{w|k}$ from all word master vertices. Note that we do not aggregate $N_k = \sum_dN_{k|d}$ since the number of documents may be 100+ times larger than word number.

\subsection{Related work in parallelization}
Recently, there are many efforts towards to build distributed topic modelling system.
They have different partition approaches and synchronization approaches. Almost all works~\cite{SparkLDA,LightLDA,F+LDA} partition the corpus in a document-wise way.
Instead, {\name} permits any kind of partition methods.
No parallelization work is found to have strictly synchronization semantic that at least $N_k$ is not synchronized.
The above works all achieve the synchronization on $N_{k|d}$ since they choose document-wise partitioning.
They further processes one document partition in a word-by-word order, and synchronization cross-machine happens once a mini-batch with certain workload is completed.
However, SparkLDA and F+LDA schedule the tasks that only one task is working on a word mini-batch that they have no conflict on $N_{w|k}$.

\subsection{Utilities}
{\name} supports both log likelihood and perplexity as the metric to evaluate the model convergence.
{\name} also supports flexible termination condition, which can be based on a given number of training iteration or the perplexity value.
Besides the core of model training, {\name} provides useful utilities in entire lifecyle.

\noindent\textbf{Incremental training.}
Model can be saved in the middle of training rather than waiting the end of the training. In this way,
users can terminate the training if they think the model is converged.
Incremental training that the model can be initialized by a pre-trained model is also supported.
This is useful when users are not satisfied the trained model and continue the model training.
The model re-training could be equipped with better hyper-parameters, or new training data, etc.

\noindent\textbf{Merge duplicated topics.}
Frequent words often dominate more than one topics, and the learned
topics are similar to each other. These similar topics are duplicated~\cite{asymLDA,peacock}.
{\name} adopts the asymmetric Dirichlet prior~\cite{asymLDA} that automatically combine similar topics into
one large topic, rather than splitting topics more uniformly by symmetric priors.
This is especially useful when the number of topic number is large ($\leq 10^6$).
Besides, we also cluster topic duplicates if their L1-distance is below a threshold. The lower L1-distance threshold means that we would
remove more duplicates from large number of topics.

\noindent\textbf{Model inference.}
{\name} also supports model inference besides model training, that {\name} inferences the topic distribution over the given a document with the same CGS process.
The inference process can use the same tricks presented in {\name} (Section~\ref{sec:cgs}).
To accelerate the model inference in online applications like search engine and online advertising systems to
predict latent semantics of new user queries in real-time (usually in milliseconds) from large number of topics,
we adopts RT-LDA~\cite{peacock} that replace the sampling operation in Equation~\ref{f:3} by the max operation,
which makes RT-LDA significantly faster than standard inference equation, but still with similar perplexity.

\section{Optimizations}
\label{s:opts}

Data-parallel system helps {\name} to simplify the programming efforts, hide the system complexity and integrate with entire training pipeline.
However, it would result in sub-optimal performance compared with customized approaches.
In this section, we describe the techniques that help {\name} to achieve comparable performance and scalability.
including approximated training, network I/O reduction and several low-level optimizations.

\subsection{Approximated training}
\label{s:approx}
As proved in asynchronized update that sampling can work on staled model state, CGS training could tolerate a certain degree of approximation.
In this section, we will present two important approximations to make better tradeoffs on efficiency and model accuracy, including sparse model initialization and ``converge'' edges exclusion from sampling.

\textbf{Sparse model initialization.}
It is well known that the execution time per iteration decreases by degrees as training makes progress,
and the first several iterations are always the performance and scalability bottleneck,
i.e., the training would succeed if it passes the first iteration.
\comment{\todo{A figure to show the time per iteration distribution.}}

Usually, CGS training is initialized by first randomly sampling a topic for each token with equal topic probability, and initiating the model state $N_{k|d}$ and $N_{w|k}$ by aggregating the topic distribution for each word and document, respectively.
However, such \texttt{random initialization} would result in relatively dense topic distribution for word, especially for hot words that occurred in most of the documents.
As a consequence, this dense word-topic distribution takes more storage, memory consumption, network I/O (step 2 in Figure~\ref{fig:workflow}) and more computing complexity.
Can we initialize the model with a sparse word-topic distribution but still achieve the similar convergence speed and accuracy?
To validate the assumption, {\name} presents two \texttt{sparse initialization} approaches that demonstrate much better performance in the first several iterations and comparable or even better convergence and accuracy (See Figure~\ref{fig:lda-earlyterm}).
\begin{enumerate}
  \item \texttt{Sparsify word-topic array.} The first approaches is to directly sparsify word-topic array ($N_{w|k}$). Assume that there are $T$ tokens of word $w$ in the corpus.
        Given total topics $K$ and the sparsity degree $deg \ll 1$, it first randomly samples $deg*K$ topic set $S$ from $K$ topics, then randomly samples a topic from $k \in S$ for each token of that word with equal probability, and updates $N_{k|d}$ and $N_{w|k}$ such that $\sum_kN_{w|k} = T$.  Such sparse initialization would largely relieve the performance burden in the first iteration.
        However, it has side effects on model accuracy. The good side is that it reduces the possibility to allocate the same topic for two words that should be with different topic, since their topic overlapping probability is reduced due to sparse initialization; on the contrary, the bad side is that this also reduces the possibility if the two words should be with the same topic.
        Our evaluation indicates that the CGS process still converges and gradually recovers the side effect of sparse initialization.
        This optimization is essentially to gradually amortize the cost of first iteration to the following iterations.
        We further neutralize the side effect by increasing the $\beta$ value in decomposed part $N_{k|d}*(N_{w|k} + \beta)$ for those topics that are not assigned during initialization.
  \item \texttt{Sparsify document-topic array.} The second approach is to sparsify document-topic array ($N_{k|d}$) with the same method, that thus indirectly results in sparse word-topic array.
\end{enumerate}

\textbf{``Converged'' token exclusion.}
We observed that different tokens are with different convergence rate. However, the CGS process still be applied normally without differentiation.
We present ``converged'' \texttt{token exclusion} that excludes tokens from CGS process that have been converged,
which will largely reduce the workload per iteration, especially as for later iterations that almost tokens are converged (See Figure~\ref{fig:lda-earlyterm-rate}).
The question is how to identify a token is converged, and how to neutralize the possible side effect?
We treat a token is converged if current sampled topic is the same as topic sampled in last iteration.
To reduce the side effect, we do not exclude the converged token directly, instead, they are still sampled with a probability.
Such probability considers how many iterations a token has not been processed ($i$) and how many times it was processed but with the same sampled topic ($t$).
Both $i$ and $t$ are zeroed for clearing once the sampled topic is different.
Thereby, the probability is $2^{i-t}$ that the probability has positive correlation with $i$ but has negative correlation with  $t$.
We also support user to configure this optimization to be enabled only after certain iterations.
\comment{\todo{overhead due to extra storage cost.}}

\subsection{Network I/O reduction via delta aggregation}
\label{s:delta}
Network I/O still matters that a significant portion of time is spend in average, especially for large scale execution with large number of partitions.
As shown in Figure~\ref{fig:workflow}, there are only four steps (except step 3) involved with network I/O, where the size of $N_k$ in step 1 is negligible, and
the size of $N_{k|d}$ and $N_{w|k}$ (step 2, 3 and 5) has already been reduced by sparse initialization.
In this section, we describe techniques to further reduce the network I/O in step 4 that each vertex slaves sends its locally aggregated $N_{k|d}$ and $N_{w|k}$ to master.
The $N_{k|d}$ to be transferred are locally aggregated from $Z_{dw}$ from all tokens of document $d$, and the same is for $N_{w|k}$.
Obviously, they are positively correlated with the number of tokens per document and per word, respectively.
With the same insight as ``converged'' token exclusion, high  proportion of tokens are converged without topic change, we present
delta aggregation that only the topic of changed tokens is aggregated in local and transferred to master.
Therefore, the network I/O would be largely reduced as the model becomes converged.
This requires to store the old topic sampled last time, other than new topic sampled currently, which doubles the attribute size in edge.
Moreover, the effectiveness would be offset by ``converged'' token exclusion.
Thereby, we will disable this optimization if token exclusion is enabled.
However, it is worth noting that unlike token exclusion, this optimization would not affect the model accuracy.

\subsection{Low-level optimizations}
\label{s:lowlevel}
Besides the design principle, the performance also lies in the detail.
In this section, we introduce several low-level optimizations that are proved to be generally effective, including efficient data structure that exploits the sparsity and redundant computing elimination.

\noindent\textbf{Sparse data structure.}
The right choice on data structure is crucial for performance. Here we discuss three different choices that exploits the inherent sparsity in word-topic array $N_{w|k}$ and document-topic array $N_{k|d}$,
including \texttt{DenseVector} and \texttt{SparseVector} that provided in MLlib, as well as our new proposed \texttt{CompactVector}.
DenseVector is represented as an array, and SparseVector is represented by an index array that records the indices of non-empty elements and an value array that records the corresponding values. For instance,
a vector $(1, 0, 0, 0, 0, 3)$ can be represented in dense format as $[1, 0, 0, 0, 3]$
in sparse format as $(6, [0, 5], [1, 3])$, where $6$ is the size of the vector.
Compared with dense vector, it is more memory efficient if vector has large sparsity,
but with increased cost for operations such as search that the complexity is increased from $O(1)$ to $O(log(length))$.
Note that it would result in more memory space for vector with less sparsity.
The tipping point is when the sparsity is 0.5 that only half of the elements are empty, where the total length of index and value array is equal to the original vector length.
Instead, we provide a new sparse vector representation, called \texttt{CompactVector}, that also includes a value array as \texttt{SparseVector} and a different index array.
The index array is composed of $(s, n)$ pairs where $s$ records the starting index of an empty sequence and $n$ records the number of non-empty elements before position $s$. For example,
$(1, 0, 0, 0, 0, 3)$ is represented as $(6, [(1, 1)], [1, 3])$.
Figure~\ref{fig:cv} describes how to get value from CompactVector, given the original index $x$.
The time complexity is $O{logN}$ where $N$ is the number of empty sequences, i.e., $N$ is the number of non-empty sequences, thus $N$ is less than the number of non-empty elements $E$, since
sequence is composed of at least one element.
Thereby, the time complexity is lower than it in SparseLDA that is with $O(logE)$
In addition, the size of CompactVector could be smaller specifically when $\frac{E}{N} \geq 2$, consider that CompactVector represents a sequence with two (both $s$ and $n$) data element.
The disadvantage of CompactVector is that the insertion is much costly with $O(N)$ complexity.

The right choice should tradeoff between the space requirement and computing cost.
Generally, \texttt{CompactVector} is more suitable for scenarios where space is critical and almost operations are read;
\texttt{SparseVector} is suitable for vectors with large sparsity; and \texttt{DenseVector} is suitable for dense vector with many write operations, since array in Scala can be updated in place while the others are immutable that a new operation is required each time the value is changed or a new value is inserted.
Take the computing of $N_{k|d}*(N_{w|k} + \beta)$ as an example, $(N_{w|k}$ is read given the topic $k$ where $N_{k|d}$ is non-zero. Such read is with $O(logK_w)$,
which increases the complexity of probability computing  from $O(K_d)$ to $O(K_d*logK_w)$, thus {\name} chooses \texttt{DenseVector} to convert $N_{w|k}$ from sparse vector.

\begin{algorithm}[t]
\begin{algorithmic}[1]
    \State{\textbf{Input:} $CV = (len, index, value)$, original index $x$.}
    \State{\textbf{Output:} the value indexed at $x$.}
    \Procedure{GetValue($CV$, $x$)}{}
        \State {$(s_i, n_i), (s_j, n_j) \gets BSearch(x, CV)$}
        \State {assert($s_i \leq x \leq s_j$)}
        \If{$x \neq s_i \; \&\& \; x \neq s_k$}
            \If {$x \geq s_j - (n_j - n_i)$}
                \State {$d = x - (s_j - (n_j - n_i))$}
                \State{\Return {$value[n_i + d]$}}
            \EndIf
        \EndIf
        \State{\Return {NULL}}
    \EndProcedure
\end{algorithmic}
\caption{Get value from CompactVector.}
\label{fig:cv}
\end{algorithm}

\noindent\textbf{Alias table.}
We use alias table, $gTable$ for $\frac{\alpha*\beta}{N_k+W\beta}$ and $wTable$ for $\frac{N_{wk}*\alpha}{N_k+W\beta}$, to avoid re-computation cost and save the sampling complexity to $O(1)$.
However, the time complexity to build alias table is $O(K)$ and $O(K_w)$, respectively. Moreover, Each word vertex has a $wTable$ , which requires more memory space.
We reduce the memory consumption by processing the tokens in word-by-word fashion that reduces the lifecycle of $wTable$ thus unused $wTable$ would be freed (GC).
To reduce the creation cost, we further refine the algorithm presented in AliasLDA~\cite{aliasLDA}.
First, we only maintain the $H$ queue that keeps the topic information ($(k, p_k)$) that is with higher probability than the average $\frac{1}{K_w}$, and do not maintain the $L$ queue described in in AliasLDA.
Instead, we directly insert the topic information with lower probability into the bin of alias table in a sequential way.
Second, when create alias table for $N_{k|d}$ (used in LightLDA), the probability (count) is integer, but the average probability would be float that is the result of dividing the sum by $K_d$.
Consequently, the split probability in a bin should be float. Instead, we first multiply $K_d$ for each individual probability, therefore, both the average and the split probability are also integer.
In this way, we avoid the costly divide operation, and simultaneously save the space.

\noindent\textbf{Redundant computing elimination.}
There are many redundant computing in CGS formula~\ref{f:3}. For instance, $\frac{1}{N_k+W\beta}$ will be used many times during entire iteration,
thus we can pre-compute it first and re-use the result later~\footnote{Note that $N_k$ is constant with value computed in last iteration.}
The following code snippet (Algorithm~\ref{alg:pre}) depicts how we decompose the computation and eliminate the redundancy.
Besides redundancy elimination, the multiplication between scalar and vector (denoted as $.*$) enables the instruction level parallelism that uses vectorization via SIMD instructions.
Lastly, this is also CPU cache friendly. For example, only $N_k$ is accessed to compute $t1$, and only $N_{w|k}$ is accessed to compute $wSparse$.
It is worth to note that such concept can also be applied to other CGS decompositions.

\comment{
\begin{figure}[t]
\vspace*{-1ex}
\begin{center}
\small{
\begin{lstlisting}[mathescape=true]
$t1$ = $\frac{1}{N_k+W\beta}$ // vector
$t2$ = $\frac{K\alpha}{N+\alpha'}$ // constant
$t3$ = $(\frac{\alpha'}{K} - W\beta)$ // constant
$t4$ = $\frac{\alpha_k}{N_k+W\beta}$ = $\frac{K\alpha \frac{N_k + \frac{\alpha'}{K}}{N+\alpha'}}{{N_k+W\beta}}$ = $\frac{K\alpha}{N+\alpha'}$*$\frac{N_k+\frac{\alpha'}{K}}{N_k+W\beta}$ = $\frac{K\alpha}{N+\alpha'}$*$(1+\frac{\frac{\alpha'}{K} - W\beta}{N_k+W\beta})$ = $\frac{K\alpha}{N+\alpha'}$ + $\frac{\frac{K\alpha}{N+\alpha'}(\frac{\alpha'}{K} - W\beta)}{N_k+W\beta}$
   = $t2$ .+ ($t2*t3$).*$t1$ // vectorization
$t5$ = $\beta$.*$t1$ // vectorization
$gDense$ = $\frac{\alpha_k*\beta}{N_k+W\beta}$ = $\beta$.*$t4$ // vectorization
(foreach $w \in W$) {
  $wSparse$ = $\frac{N_{w|k}*\alpha_k}{N_k+W\beta}$ = $N_{w|k}$*$t4(k)$ // $N_{w|k} \neq 0$
  $t6$ = $\frac{N_{w|k}+\beta}{N_k+W\beta}$ = $t5$ + $N_{w|k}$*$t1(k)$ // $N_{w|k} \neq 0$
  (foreach $d \in D$) {
    $dSparse$ = $N_{k|d}(\frac{N_{w|k}+\beta}{N_k+W\beta})$ = $N_{k|d}$*$t6(k)$ // $N_{k|d} \neq 0$
  }
}
\end{lstlisting}
}
\vspace{-0.4cm}
\caption{Redundant computing elimination.}
\label{fig:pre}
\end{center}
\vspace*{-3ex}
\end{figure}
}

\begin{algorithm}[t]
\begin{algorithmic}[1]
\State{$t1$ = $\frac{1}{N_k+W\beta}$}
\State{$t2$ = $\frac{K\alpha}{N+\alpha'}$}
\State{$t3$ = $(\frac{\alpha'}{K} - W\beta)$ }
\State{$t4$ = $\frac{\alpha_k}{N_k+W\beta}$ = $\frac{K\alpha \frac{N_k + \frac{\alpha'}{K}}{N+\alpha'}}{{N_k+W\beta}}$ \newline
  = $\frac{K\alpha}{N+\alpha'}$*$\frac{N_k+\frac{\alpha'}{K}}{N_k+W\beta}$ = $\frac{K\alpha}{N+\alpha'}$*$(1+\frac{\frac{\alpha'}{K} - W\beta}{N_k+W\beta})$ \newline
  = $\frac{K\alpha}{N+\alpha'}$ + $\frac{\frac{K\alpha}{N+\alpha'}(\frac{\alpha'}{K} - W\beta)}{N_k+W\beta}$ \newline
  = $t2$ .+ ($t2*t3$).*$t1$ }
\State{$t5$ = $\beta$.*$t1$}
\State{$gDense$ = $\frac{\alpha_k*\beta}{N_k+W\beta}$ = $\beta$.*$t4$}
\For{each $w \in W$}
  \State{$wSparse$ = $\frac{N_{w|k}*\alpha_k}{N_k+W\beta}$ = $N_{w|k}$*$t4(k)$ // $N_{w|k} \neq 0$}
  \State{$t6$ = $\frac{N_{w|k}+\beta}{N_k+W\beta}$ = $t5$ + $N_{w|k}$*$t1(k)$ // $N_{w|k} \neq 0$}
  \For{each $d \in D$}
    \State{$dSparse$ = $N_{k|d}(\frac{N_{w|k}+\beta}{N_k+W\beta})$ = $N_{k|d}$*$t6(k)$ // $N_{k|d} \neq 0$}
  \EndFor
\EndFor
\end{algorithmic}
\caption{Redundant computing elimination.}
\label{alg:pre}
\end{algorithm}

\noindent\textbf{Others.}
{\name} also implements several other optimizations that are generally beneficial.
First, {\name} tries to reuse the same generated random number as far as possible to avoid cost of random number generation.
For example, there are three random number generations, the first is used to locate $gTable$, $wTable$ or $gSparse$, the second is for locate the specific bin in alias table, and the
last one is to locate the high or low region in a bin. Apparently, the last two can use the same random number.
Second, we can pre-generate $n$ random numbers for $n$ different tokens in the same $(d, w)$ pair, in this way, the CDF sampling cost is reduced from $O(n*logK_d)$ to $logn*logK_d$ that only $logn$ passes of CDF binary search are required.
Lastly, {\name} also implements optimization to exploit the difference between hot and long-tail word, as described in LightLDA~\cite{LightLDA}

\section{Implementation}
\label{s:impl}
We encountered scalability issues due to the inherent inefficiency of managed language (Scala) and framework cost of GraphX.
The implementation must balance resource (CPU, network and memory) usage that no resource is the bottleneck and all are fully utilized.

\noindent\textbf{Memory.}
Memory is the major bottleneck to when we first try to scale out {\name}. Data-parallel system like Spark is designed to process one partition per core and the whole partition must be loaded in memory.
``Out of memory'' occurs frequently if many partitions (we have 16-32 cores per machine) loaded at the same time.
The dilemma lies in that if too many partitions would reduce the memory consumption but with the increased network I/O.
We observed that these partitions in a machine may share common data, such as the same word or document may exist in multiple partitions, thus the same as the corresponding word-topic or document-topic array.
Instead, we load less partitions at one time and use multi-thread computing in a partition to fully utilize the CPU cores.
More specifically, edges is sorted queued in word-by-word order in a partition (already done by GraphX), and work-stealing with word granularity is adopted among multiple threads that once a thread completed all edges of one word, it will get all edges of the first word from edge queue. This achieve relative good load balance. The more fine-grained edge granularity is also feasible that the edges of the same word are processed in parallel, where
$N_{w|k}$ is further shared among threads.
Besides CGS processing (step 3 in Figure~\ref{fig:workflow}), we re-implement some GraphX APIs (except shuffling operator) to make them multi-threaded,
such as \texttt{ShipVertexAttributes} (step 2) and \texttt{aggregateMessages} (step 4).
Actually, we abandon \texttt{aggregateMessages} that updates vertex attribute with value aggregated from edges,
and constraints that the edge attribute type must be the same as with vertex attribute, thus costly type conversion happened.
Instead, we directly operate on the $Graph$ data structures (e.g., edge array and vertex index array in EdgePartition, vertex array and routing table in VertexPartition).
This will significantly reduce memory costs and scale 10X up than original GraphX implementation.

Many GraphX APIs would create many intermediate objects that raise higher memory costs and the overhead of garbage collection.
For example, to represent the bin of alias table ($(i,h,p_h)$), we use three arrays with primitive type instead of one array with Scala \texttt{Tuple} to avoid the boxing/unboxing overhead.

\noindent\textbf{CPU.}
Once we fixed the memory limitation via multi-threading, we found that high CPU cost on RDD decompression and deserialization that is processed by a single thread.
Therefore, we prefer to configure RDD in uncompressed and deserialized format.
Besides the optimizations presented in Section~\ref{s:lowlevel}, the avoid of boxing/unboxing and generation of closures can also reduce the CPU cost.

\noindent\textbf{Network.}
The relax on memory burden can enable more partitions, but this will increase the network I/O. We adopt Kryo serialization library in Spark
that is significantly faster and more compact than default Java serialization.
Shuffling cost is largely reduced when Kryo is enabled for serializing shuffling data.

\section{Evaluation}
\label{s:eval}
This section describes the evaluation to demonstrate the effectiveness and efficiency of {\name}.

\subsection{Evaluation design}

\begin{table}
\centering
\begin{tabular}{|l|r|r|r|l|}
\hline
Dateset      & Tokens       & Words    & Docs       & $T/D$ \\ \hline
NYTimes      & 99,542,125     & 101,636   & 299,752     & 332              \\ \hline
BingWebC1Mon & 3,150,765,984   & 302,098   & 16,422,424   & 192              \\ \hline
BingWebC320G & 54,059,670,863  & 4,780,428  & 406,038,204  & 133              \\ \hline
\end{tabular}
\caption{Three different datasets used in evaluation.}
\label{tab:dataset}
\end{table}

\noindent\textbf{Datasets.}
We use 3 different datasets, including a small sized NYTimes~\cite{nyt} (about 520MB), a medium sized one month web chunk data indexed by Bing News (about 17GB),
and a large scale Bing web chunk data (320G). They are all pre-processed and saved as \texttt{libsvm} format.
The detailed information is listed in Table \ref{tab:dataset}.

\noindent\textbf{Evaluation design.}
The evaluation aims to evaluate: 1). the algorithm effectiveness and efficiency in {\name} compared with LightLDA who represents the state of art.
2). the scalability of {\name} that varies topic number, dataset size and number of machines.
3). the effectiveness of proposed techniques in (\name).

\noindent\textbf{Cluster configuration.}
We have two Spark clusters with different scale. The small one is in lab environment and has 10 homogeneous computing nodes are connected via 40Gbps
Infiniband network and each node has 16 2.40GHz Intel(R) Xeon(R) CPU E5-2665 cores and 128GB memory. There are
1 driver configured with 5GB memory and 10 workers configured with 100GB memory.
The experiments against NYTimes and BingWebC1Mon are conducted in this small cluster, where NYTimes is partitioned into
20 partitions and each partition has 8 threads, BingWebC1Mon has 80 partitions and each one has 2 threads.
The large Spark cluster is deployed on a multi-tenancy data center managed by Yarn~\cite{yarn} that the resource is not always guaranteed.
An executor is configured to have 20GB memory and 14 cores.
The scalability experiments against BingWebC320G are conducted in this cluster.

\subsection{Effectiveness and efficiency of CGS algorithm  in {\name}}

\begin{figure}[t]
\centering
\includegraphics[width=0.5\textwidth]{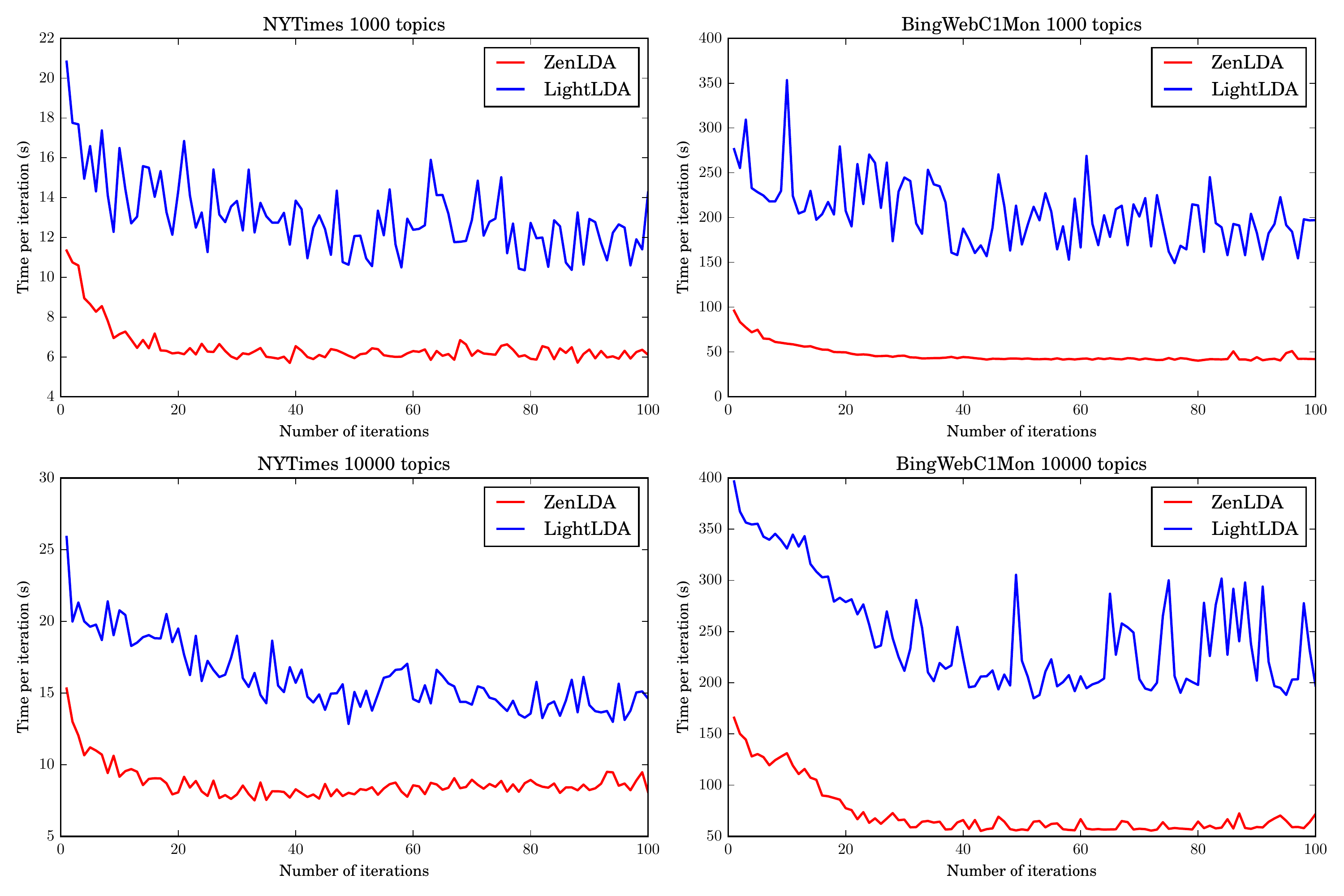}
\caption{Execution time comparison between ZenLDA and LightLDA.}
\label{fig:lda-algo-time}
\end{figure}

\begin{figure}[t]
\centering
\includegraphics[width=0.5\textwidth]{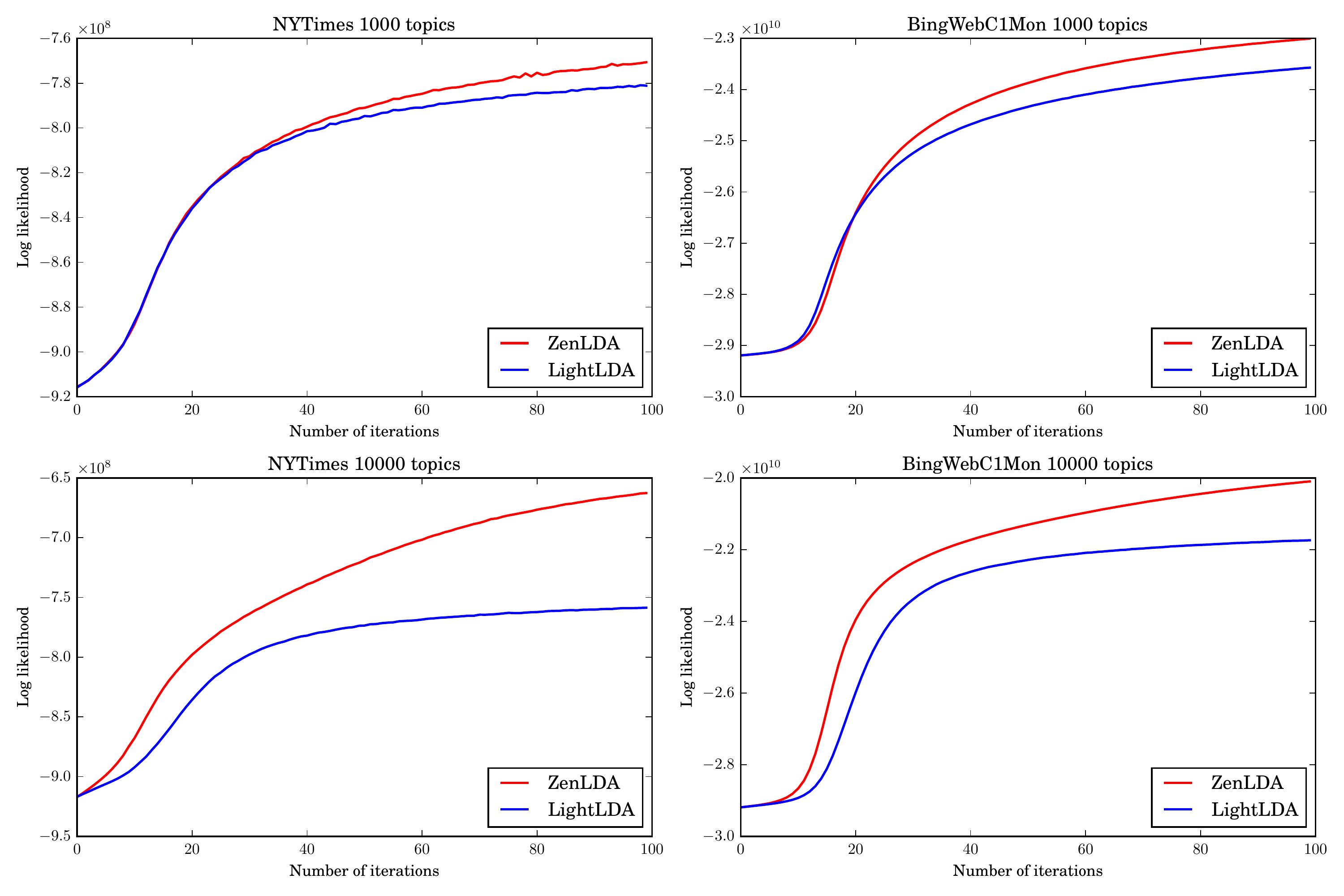}
\caption{Log-likelihood comparison between ZenLDA and LightLDA.}
\label{fig:lda-algo-pplx}
\end{figure}
To compared {\name} with LightLDA, we also implemented it with the same framework with 8 Metropolis-Hasting steps.
We excluded sparse initialization and toke exclusion, and applied the same optimizations described in Section~\ref{s:opts} to LightLDA, and the only difference is the algorithm.
The comparison is against NYTimes and BingWebC1Mon datasets with 1,000 and 10,000 topics, respectively.  Both $\alpha$ and $\beta$ are 0.01.
Both execution time per iteration and log-likelihood per iteration are compared, and the result is shown in Figure \ref{fig:lda-algo-time} and Figure \ref{fig:lda-algo-pplx}, respectively.
Note that it excludes the log-likelihood computing time, and the spikes in Figure \ref{fig:lda-algo-time} stems from full GC in JVM.
We can see significant execution reduction (2-6X speedup) and better model accuracy than LightLDA for all experiments.
Different datasets has different speedup, the larger the dataset is, the more speedup achieved (2X in NYTimes and about 4-6X in BingWebC1Mon).
There is no obvious speedup difference when topics varies from $1,000$ to $10,000$.
The experiments also show that as the number of topics increased, the performance is still almost the same in $\name$ but increases a little (about from 13s to 17s in NYTimes and from 220s to 250s) in LightLDA.
Performance is slowdown with sub-linearity if dataset is increased.

The performance result is a little bit ``surprising'', consider that LightLDA has $O(1)$ perplexity.
As discussed in Section~\ref{s:relcgs},
the MH-step in LightLDA would be more costly due to the computation of true probability, which requires $O(max(logK_w,logK_d)$ complexity since $N_{k|d}$ and $N_{w|k}$ in our implementation is sparse thus with $O(logK_w)$ or $O(logK_d)$ complexity to read the value. And there are $\#MH$ (8 in our implementation) MH-steps. As a comparison, the complexity in {\name} can be as low as $min(K_d, K_w)$.
With respect to log-likelihood, {\name} outperforms LightLDA, and even more significant as the number of topics increased.
This may be due to facts that the asymmetric prior is used in {\name} and the proposal distribution in LightLDA is an approximation of the true probability.
~\footnote{Note that we cannot directly compare the result with Figure 13 and 14 in LightLDA paper, since we double confirmed with LightLDA author that they used different log likelihood formula we used
($llh = \sum_w{log\sum_k{\frac{N_{k|d}+\alpha_k}{N_d+K\alpha_k}*\frac{N_{w|k}+\beta_w}{N_k+W\beta_w}}}, \alpha_k = \frac{N_k+\alpha'}{N+K\alpha'}$).}

We also compared with other algorithms such as SparseLDA (we implemented in the same framework), and the EM based implementation in MLLib~\footnote{We cannot compare SparkLDA since it is not open-sourced. But we believe {\name} will win since SparkLDA uses standard CGS algorithm.}.
SpasreLDA is much slower that we did not run full length of SparseLDA but only with the first 15 iterations. It spends about 27,707 seconds (10,000) while {\name} only needs 1,907 seconds.
And the EM algorithm in MLlib even cannot finish the first iteration against BingWebC1Mon dataset with errors reported 1 hour later.

\subsection{Scalability}

\begin{figure}[t]
\centering
\includegraphics[width=0.4\textwidth]{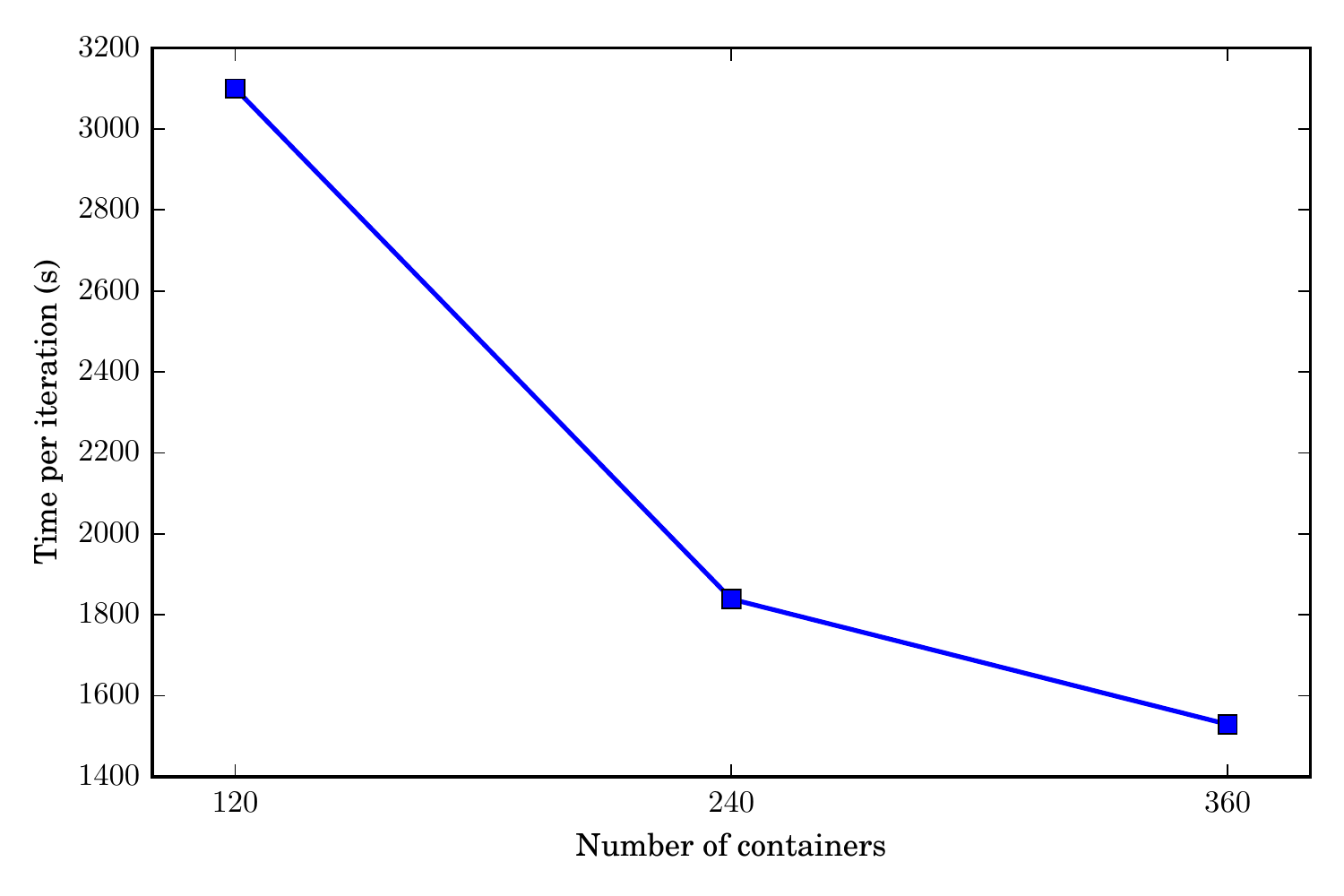}
\caption{Execution time change curve as executor number varies.}
\label{fig:lda-hemera-time}
\end{figure}

The scalability experiments are conducted against the largest dataset and run on our multi-tenancy data center.
Figure \ref{fig:lda-hemera-time} indicates that {\name} can support super large dataset in acceptable time.
When 2X more executors (containers in Yarn) joined in (240 VS 120), the performance is almost linearly speedup.
As we continue to add more executors (360), the performance can still be improved, but with less speedup due to the network I/O becomes larger.

\begin{figure}[t]
\centering
\includegraphics[width=0.4\textwidth]{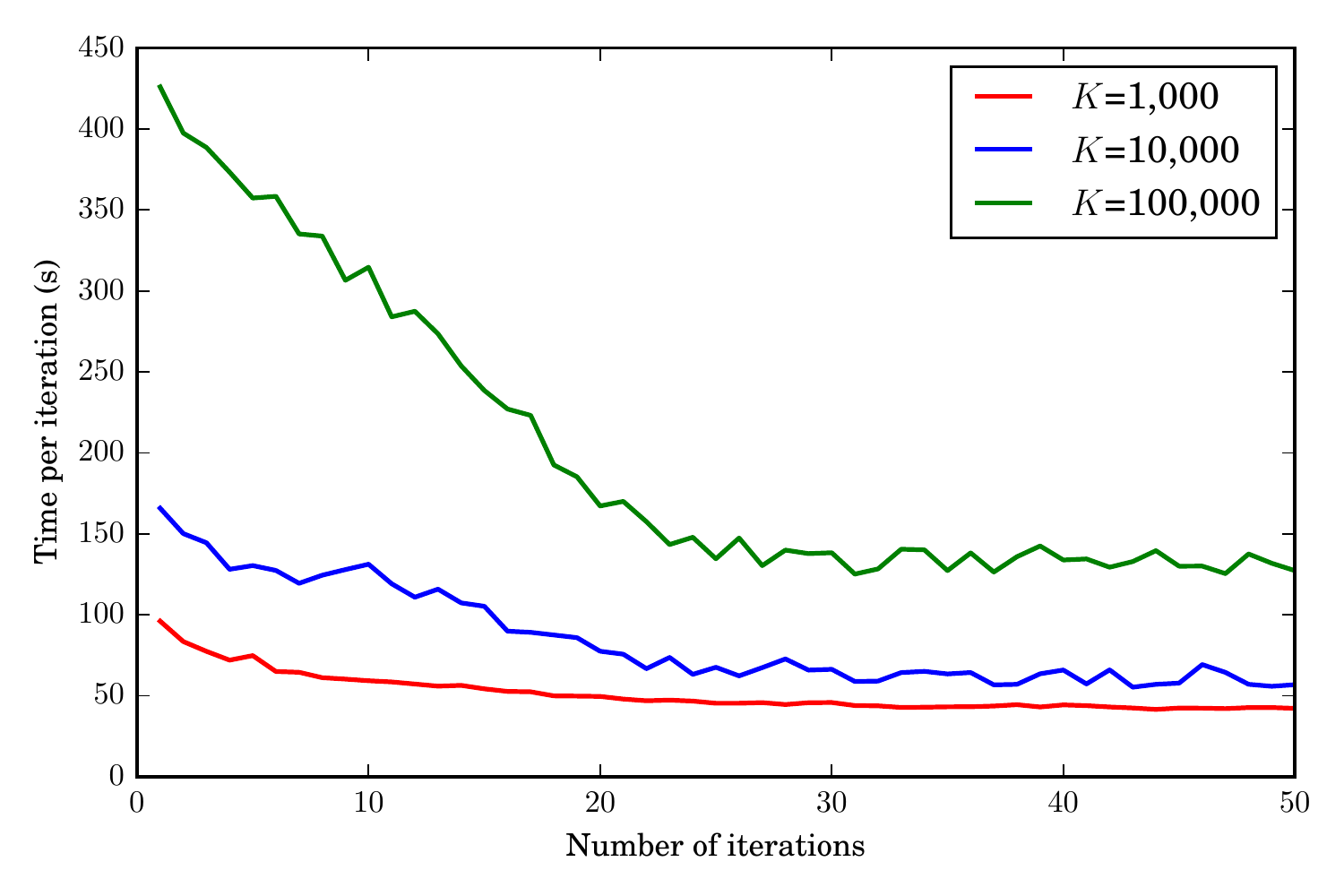}
\caption{Execution time change curve as topic number varies.}
\label{fig:lda-topics-time}
\end{figure}

We also evaluated the performance when topic number varies.
The experiment is conducted against BingWebC1Mon with 1,000, 10,000 and 100,000 topics, respectively.
Their training time of first 50 iterations is shown in Figure \ref{fig:lda-topics-time}.
When $K=10,000$, the average time per iteration is only increased a little bit compared with $K=1,000$.
Even with 100X more topics ($K=100,000$), the time is only increased by about 3X.

\subsection{Optimization evaluation}
%This section evaluates the effectiveness of optimizations presented in Section~\ref{s:opts}

\begin{figure*}[t]
\centering
\includegraphics[width=\textwidth]{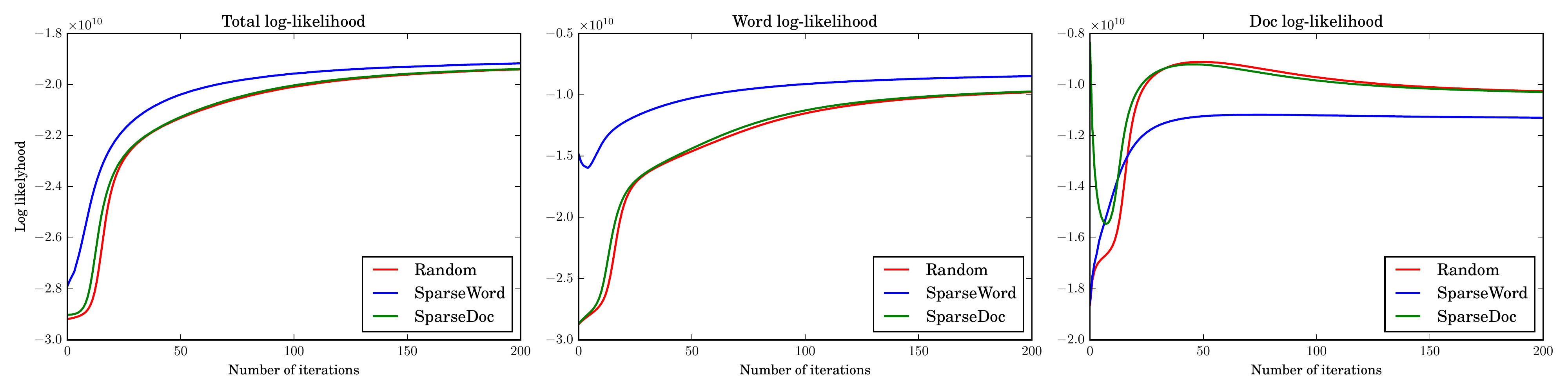}
\caption{Log-likelihood comparison among different initializations.}
\label{fig:lda-init-pplx}
\end{figure*}

\begin{figure}[t]
\centering
\includegraphics[width=0.4\textwidth]{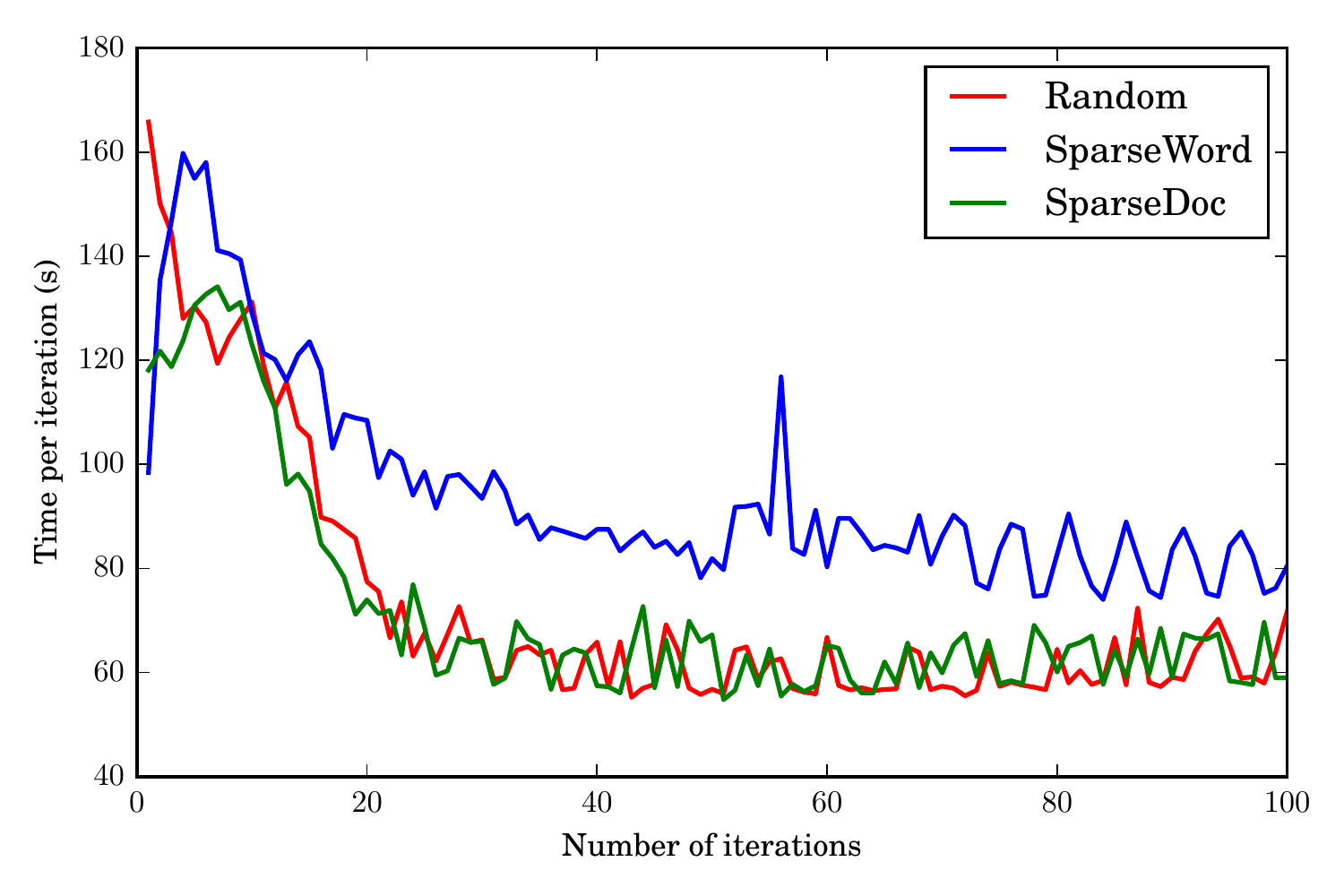}
\caption{Sampling time comparison among different initializations.}
\label{fig:lda-init-time}
\end{figure}

\begin{figure*}[t]
    \centering
    \begin{subfigure}{0.32\textwidth}
    \includegraphics[width=\linewidth]{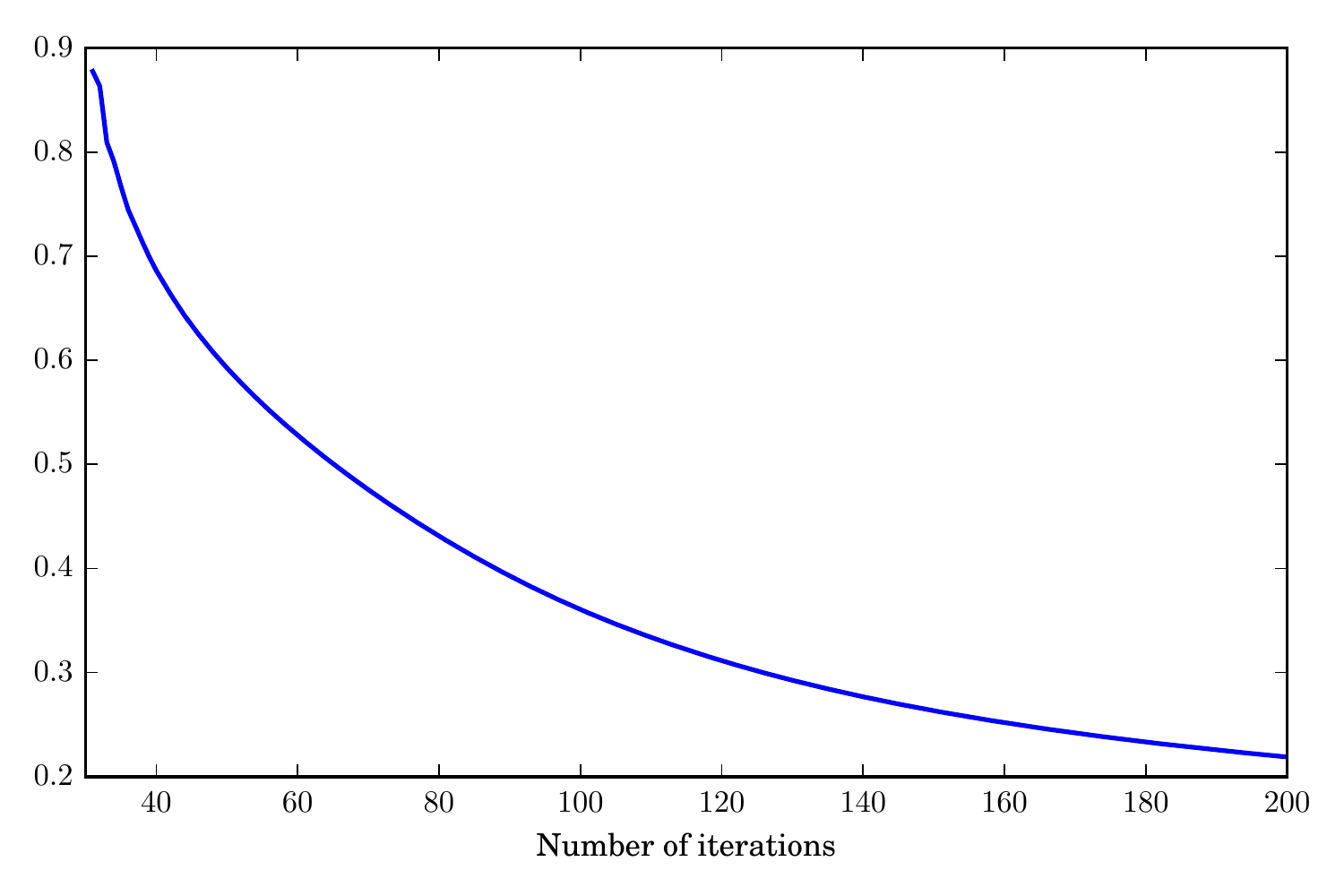}
    \label{fig:lda-earlyterm-rate}
    \end{subfigure}
    \hspace*{\fill}
    \begin{subfigure}{0.32\textwidth}
    \includegraphics[width=\linewidth]{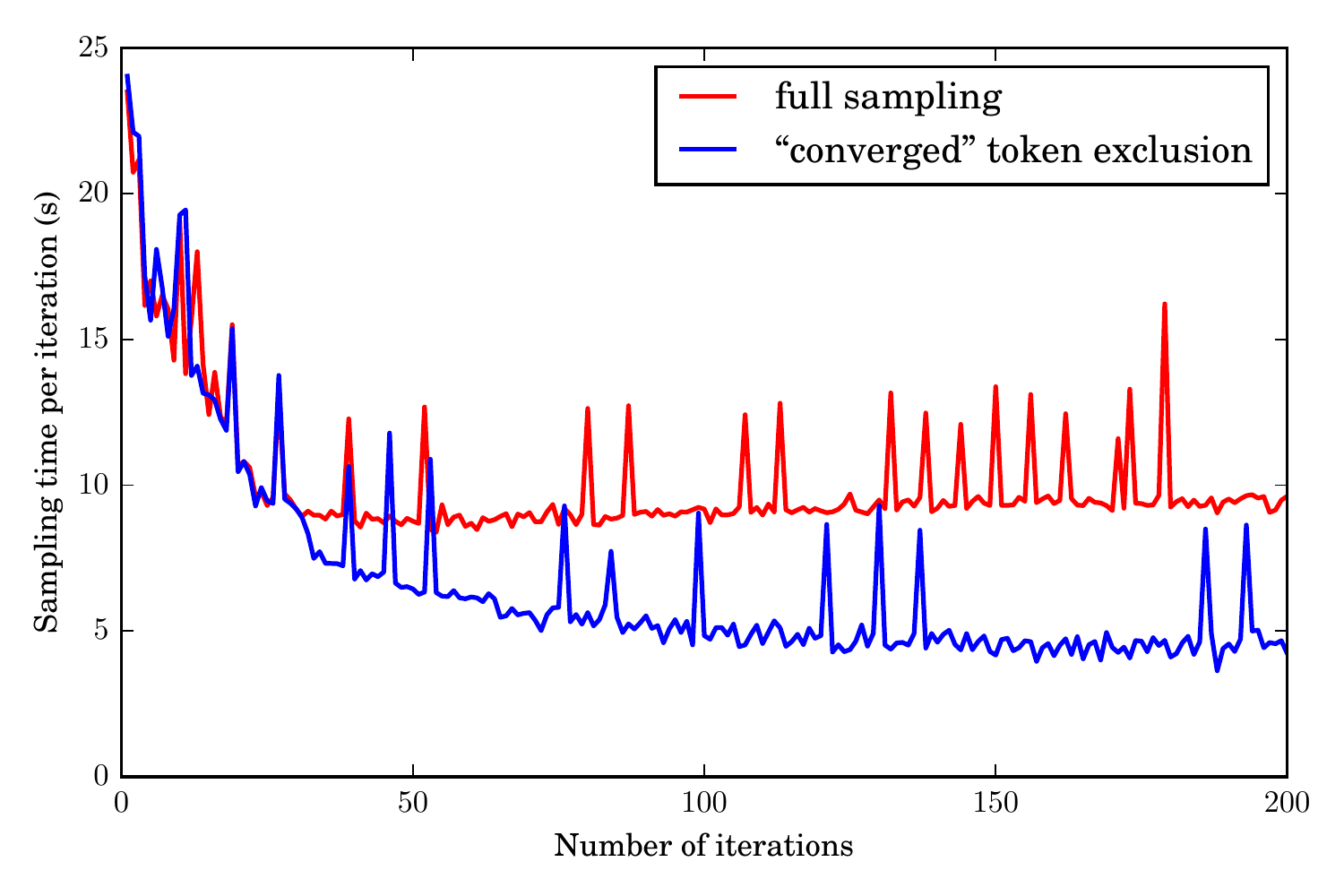}
    \label{fig:lda-earlyterm-time}
    \end{subfigure}
    \hspace*{\fill}
    \begin{subfigure}{0.32\textwidth}
    \includegraphics[width=\linewidth]{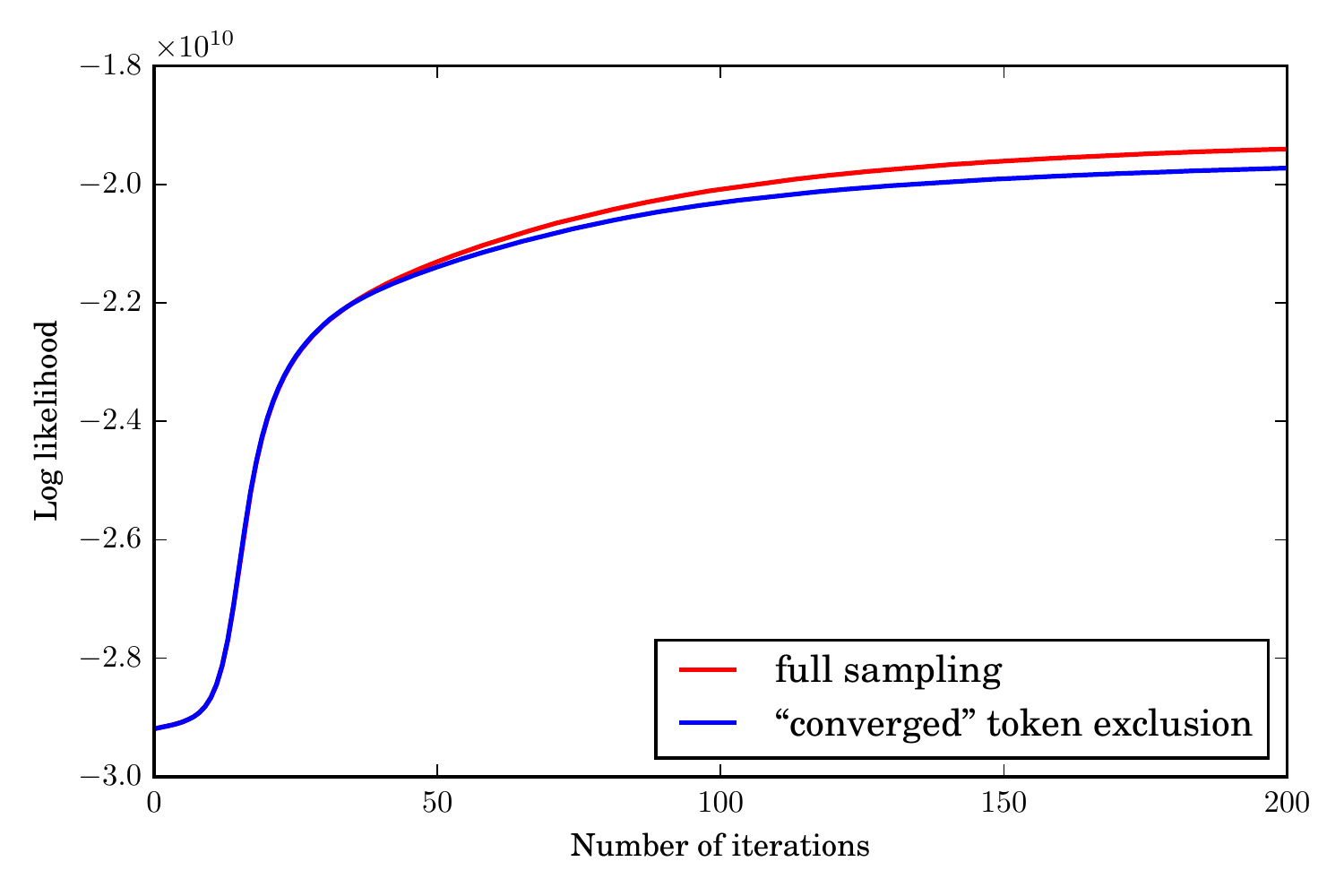}
    \label{fig:lda-earlyterm-pplx}
    \end{subfigure}
    \caption{(a) Change rate of token's topic assignments.
        (b) Sampling time with or without ``converged'' token exclusion.
        (c) Log-likelihood with or without ``converged'' token exclusion.}
    \label{fig:lda-earlyterm}
\end{figure*}

\noindent\textbf{Sparse initialization.}
Figure~\ref{fig:lda-init-pplx} shows the log-likelihood of different initialization strategies.
Specially, we further split log-likelihood into word log-likelihood and doc log-likelihood.
With respect to accuracy, sparse initialization of word-topic distribution (\texttt{SparseWord}) can even achieve better total and word log-likelihood, but with worse document log-likelihood.
In contrast, sparsifying document-topic distribution(\texttt{SparseDoc}) only achieves better doc log-likelihood in the first several iterations, and
is ``dragged'' to normal distribution as random initialization.
With respect to performance, Figure~\ref{fig:lda-init-time} shows that both SparseWord and SparseDoc make the sampling time faster than random initialization
at the first several iterations. This is helpful to reduce the scalability bottleneck.
However, it gradually increases to normal performance as random initialization as we expected, and even higher in SparseWord because of the increased $K_d$ (the worse document log-likelihood, the dense document-topic distribution).

\noindent\textbf{``Converged" token exclusion.}
{\name} chooses to turn on this optimization after the 30th iteration.
Both sampling time (exclude time on shuffling) and log-likelihood are compared.
The result shown in Figure \ref{fig:lda-earlyterm-time} and Figure \ref{fig:lda-earlyterm-pplx} indicates that ``converged" token exclusion technics can achieve about
50\% faster in later iterations, without hurting the log-likelihood much.
Figure \ref{fig:lda-earlyterm-rate} explains the underlying reason that the changing rate of topic assignment decreases as the iteration increases, with only about 22\% remained at the end.
This figures also demonstrates that delta aggregation (Section~\ref{s:delta}) can largely reduce the network I/O.
The speedup is not strictly align with the change rate since the sample rate also considers the other factors.

\noindent\textbf{Redundant computing elimination.}
We only evaluate the effectiveness of ``redundant computing elimination'' and skip other low-level optimizations that are hard to separated out.
The result in Figure \ref{fig:lda-nocc-time} shows that the sampling is faster with about 11\% improvements.
\begin{figure}[t]
\centering
\includegraphics[width=0.4\textwidth]{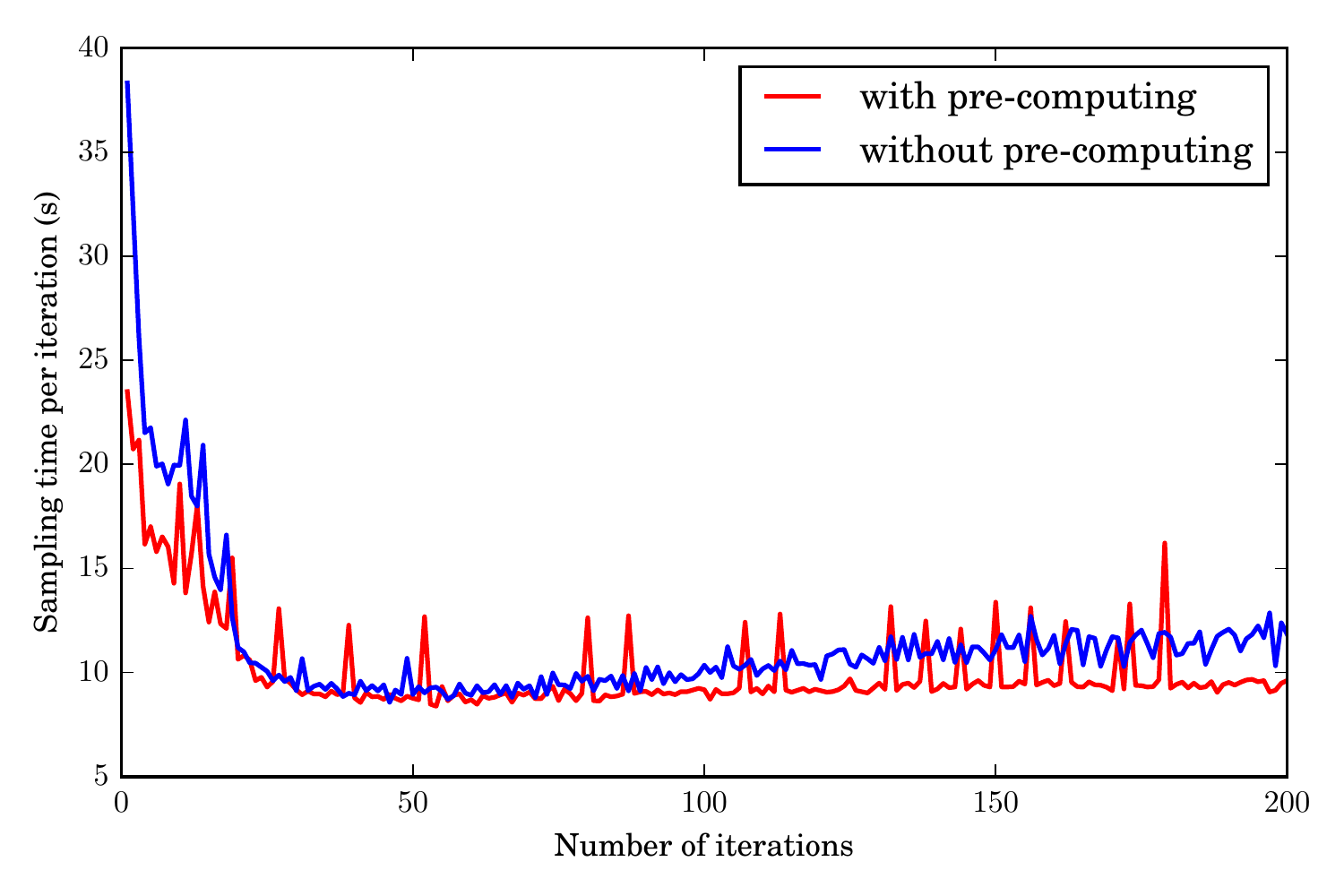}
\caption{Sampling time with or without redundant computing elimination.}
\label{fig:lda-nocc-time}
\end{figure}

\section{Discussion and Future Work}
\label{s:disc}
This section describes several points that has crucial impacts on model accuracy and system performance, and they are out of the scope of this paper and remain future work.

\noindent\textbf{Hyper-Parameter tuning.}
There are three hyper-parameters, Dirichlet priors ($\alpha$ and $\beta$), and the number of topics ($K$).
They will affect the perplexity, and how to get best configuration is still an art.
First, in {\name}  we tried asymmetric prior topic specific $\beta_k$ to offset the side effect of sparse initialization, like asymmetric prior $\alpha_k$ that aims to improve the model robustness.
However, the impact of asymmetric priors that are word or document specific and how to set the proper asymmetric prior are still unknown and remain future work.
Second, {\name} prefers larger topics at first and deduplicates topics by merging similar topics. However, too large number of topics than needed would result in the inefficient statistical inference~\cite{icml14}.
Lastly, like many other systems, there are several heuristics involved in {\name}, such as
how to set the sparsity in initialization, how to set the right sampling rate for ``converged'' token exclusion, how to dynamically enable them, as well as how to identify them and how to set the certain threshold is still manual work.

\noindent\textbf{Graph partitioning.}
Conventional graph partitioning algorithms usually assume that the network I/O introduced by cutting different vertices is the same.
However, this assumption does not remain true in LDA training. For example,
Given a vertex with less degrees but more dense word-topic and document-topic distribution, the partition strategy that cuts that vertex may introduce more the network I/O in step 2 and 4 of {\name} (Figure~\ref{fig:workflow}).
Besides, consider the situation where ``converged'' edge exclusion is enabled, the number of active edges would gradually decrease. This will conversely affect the partitioning approach that
an approach may get good load balance at first, but it gradually becomes imbalance as training proceeds.

\noindent\textbf{Others.}
Currently, GraphX does not permit different typed attributes for different vertices. This prohibits us to efficiently exploit the difference between hot words and long tail words.
It would also be interesting to theoretically analyze the impact of sparse initialization and ``converged'' token exclusion, and how to systematically neutralize their side effect remains future work.

\section{Conclusion}
\label{s:con}

In this paper, we present {\name} that proves to be an efficient and scalable collapsed Gibbs sampling system for LDA model on distributed data-parallel platform.
This reflects our belief that build distributed machine learning system is not only feasible and beneficial, but also efficient and scalable.
{\name} comes from combined innovations from both algorithm side and system side, and both are indispensable to achieve the goal.
A clear abstraction like RDD in Spark can accelerate the research on both sides, respectively.
We will continue this methodology and add more and more models in the future.

\clearpage

\bibliographystyle{abbrv}
\bibliography{reference}  %

\begin{thebibliography}{10}

\bibitem{yahooLDA}
A.~Ahmed, M.~Aly, J.~Gonzalez, S.~Narayanamurthy, and A.~J. Smola.
\newblock Scalable inference in latent variable models.
\newblock WSDM, pages 123--132, 2012.

\bibitem{mahout}
Apache.
\newblock What is apache mahout?
\newblock \url{http://mahout.apache.org/}, 2015.

\bibitem{openmp}
O.~ARB.
\newblock Openmp specifications.
\newblock \url{http://openmp.org/wp/openmp-specifications/}, 2013.

\bibitem{lda}
D.~M. Blei, A.~Y. Ng, and M.~I. Jordan.
\newblock Latent dirichlet allocation.
\newblock {\em JMLR}, 3:993--1022, 2003.

\bibitem{mllibLDA}
J.~Bradley.
\newblock Topic modeling with lda: Mllib meets graphx.
\newblock
  \url{https://databricks.com/blog/2015/03/25/topic-modeling-with-lda-mllib-meets-graphx.html},
  2015.

\bibitem{PowerLyra}
R.~Chen, J.~Shi, Y.~Chen, and H.~Chen.
\newblock Power{L}yra: Differentiated graph computation and partitioning on
  skewed graphs.
\newblock EuroSys, pages 1:1--1:15, 2015.

\bibitem{nyt}
T.~N. Y.~T. Company.
\newblock Linked open data.
\newblock \url{http://data.nytimes.com/}, 2015.

\bibitem{mapreduce}
J.~Dean and S.~Ghemawat.
\newblock {MapReduce}: Simplified data processing on large clusters.
\newblock OSDI, pages 10--10, 2004.

\bibitem{website:Giraph}
Facebook.
\newblock Large-scale graph partitioning with apache giraph, April 2014.

\bibitem{powergraph}
J.~E. Gonzalez, Y.~Low, H.~Gu, D.~Bickson, and C.~Guestrin.
\newblock Power{G}raph: Distributed graph-parallel computation on natural
  graphs.
\newblock In {\em OSDI 12}, pages 17--30, 2012.

\bibitem{mcmc}
T.~L. Griffiths and M.~Steyvers.
\newblock Finding scientific topics.
\newblock {\em PNAS}, 101:5228--5235, 2004.

\bibitem{ssp}
Q.~Ho, J.~Cipar, H.~Cui, S.~Lee, J.~K. Kim, P.~B. Gibbons, G.~A. Gibson,
  G.~Ganger, and E.~P. Xing.
\newblock More effective distributed ml via a stale synchronous parallel
  parameter server.
\newblock In {\em NIPS 26}. 2013.

\bibitem{plsa}
T.~Hofmann.
\newblock Probabilistic latent semantic analysis.
\newblock In {\em UAI}, pages 289--296, 1999.

\bibitem{dryad}
M.~Isard, M.~Budiu, Y.~Yu, A.~Birrell, and D.~Fetterly.
\newblock Dryad: Distributed data-parallel programs from sequential building
  blocks.
\newblock EuroSys, pages 59--72, 2007.

\bibitem{aliasLDA}
A.~Q. Li, A.~Ahmed, S.~Ravi, and A.~J. Smola.
\newblock Reducing the sampling complexity of topic models.
\newblock KDD, pages 891--900, 2014.

\bibitem{mu}
M.~Li, D.~G. Andersen, J.~W. Park, A.~J. Smola, A.~Ahmed, V.~Josifovski,
  J.~Long, E.~J. Shekita, and B.-Y. Su.
\newblock Scaling distributed machine learning with the parameter server.
\newblock In {\em OSDI}, pages 583--598, 2014.

\bibitem{plda+}
Z.~Liu, Y.~Zhang, E.~Y. Chang, and M.~Sun.
\newblock Plda+: Parallel latent dirichlet allocation with data placement and
  pipeline processing.
\newblock {\em ACM Transactions on Intelligent Systems and Technology (TIST)},
  2(3):26, 2011.

\bibitem{mllib}
X.~Meng, J.~K. Bradley, B.~Yavuz, E.~R. Sparks, S.~Venkataraman, D.~Liu,
  J.~Freeman, D.~B. Tsai, M.~Amde, S.~Owen, D.~Xin, R.~Xin, M.~J. Franklin,
  R.~Zadeh, M.~Zaharia, and A.~Talwalkar.
\newblock Mllib: Machine learning in apache spark.
\newblock {\em CoRR}, 2015.

\bibitem{adLDA}
D.~Newman, A.~Asuncion, P.~Smyth, and M.~Welling.
\newblock Distributed algorithms for topic models.
\newblock {\em JMLR}, pages 1801--1828, 2009.

\bibitem{FastLDA}
I.~Porteous, A.~Asuncion, D.~Newman, P.~Smyth, A.~Ihler, and M.~Welling.
\newblock Fast collapsed gibbs sampling for latent dirichlet allocation.
\newblock In {\em KDD}, pages 569--577, 2008.

\bibitem{SparkLDA}
Z.~Qiu, B.~Wu, B.~Wang, and L.~Yu.
\newblock Gibbs collapsed sampling for latent dirichlet allocation on spark.
\newblock volume~36, pages 17--28, 2014.

\bibitem{icml14}
J.~Tang, Z.~Meng, X.~Nguyen, Q.~Mei, and M.~Zhang.
\newblock Understanding the limiting factors of topic modeling via posterior
  contraction analysis.
\newblock In {\em ICML}, pages 190--198, 2014.

\bibitem{japan}
S.~Tora and K.~Eguchi.
\newblock {MPI/OpenMP} hybrid parallel inference for latent dirichlet
  allocation.
\newblock LDMTA, pages 5:1--5:7, 2011.

\bibitem{yarn}
V.~K. Vavilapalli, A.~C. Murthy, C.~Douglas, S.~Agarwal, M.~Konar, R.~Evans,
  T.~Graves, J.~Lowe, H.~Shah, S.~Seth, B.~Saha, C.~Curino, O.~O'Malley,
  S.~Radia, B.~Reed, and E.~Baldeschwieler.
\newblock Apache {Hadoop} {YARN}: Yet another resource negotiator.
\newblock SOCC, pages 5:1--5:16, 2013.

\bibitem{asymLDA}
H.~M. Wallach, D.~M. Mimno, and A.~McCallum.
\newblock Rethinking {LDA}: Why priors matter.
\newblock In {\em NIPS 22}, 2009.

\bibitem{PLDA}
Y.~Wang, H.~Bai, M.~Stanton, W.-Y. Chen, and E.~Y. Chang.
\newblock {PLDA}: Parallel latent dirichlet allocation for large-scale
  applications.
\newblock AAIM, pages 301--314, 2009.

\bibitem{peacock}
Y.~Wang, X.~Zhao, Z.~Sun, H.~Yan, L.~Wang, Z.~Jin, L.~Wang, Y.~Gao, C.~Law, and
  J.~Zeng.
\newblock Peacock: Learning long-tail topic features for industrial
  applications.
\newblock {\em ACM Trans. Intell. Syst. Technol.}, 6:47:1--47:23, 2015.

\bibitem{mpi}
Wikipedia.
\newblock Message passing interface.
\newblock \url{https://en.wikipedia.org/wiki/Message_Passing_Interface}, 2011.

\bibitem{dbh}
C.~Xie, L.~Yan, W.~jun Li, and Z.~Zhang.
\newblock Distributed power-law graph computing: Theoretical and empirical
  analysis.
\newblock In {\em NIPS 27}, pages 1673--1681.

\bibitem{petuum}
E.~P. Xing, Q.~Ho, W.~Dai, J.~K. Kim, J.~Wei, S.~Lee, X.~Zheng, P.~Xie,
  A.~Kumar, and Y.~Yu.
\newblock Petuum: {A} new platform for distributed machine learning on big
  data.
\newblock In {\em KDD}, pages 1335--1344, 2015.

\bibitem{sparseLDA}
L.~Yao, D.~Mimno, and A.~McCallum.
\newblock Efficient methods for topic model inference on streaming document
  collections.
\newblock KDD, pages 937--946, 2009.

\bibitem{F+LDA}
H.-F. Yu, C.-J. Hsieh, H.~Yun, S.~Vishwanathan, and I.~S. Dhillon.
\newblock A scalable asynchronous distributed algorithm for topic modeling.
\newblock {WWW}, pages 1340--1350, 2015.

\bibitem{LightLDA}
J.~Yuan, F.~Gao, Q.~Ho, W.~Dai, J.~Wei, X.~Zheng, E.~P. Xing, T.~Liu, and
  W.~Ma.
\newblock Light{LDA}: Big topic models on modest computer clusters.
\newblock In {\em {WWW}}, pages 1351--1361, 2015.

\bibitem{spark}
M.~Zaharia, M.~Chowdhury, T.~Das, A.~Dave, J.~Ma, M.~McCauly, M.~J. Franklin,
  S.~Shenker, and I.~Stoica.
\newblock Resilient distributed datasets: A fault-tolerant abstraction for
  in-memory cluster computing.
\newblock In {\em NSDI}, pages 15--28, 2012.

\end{thebibliography}

%\balancecolumns

%\appendix

%\balancecolumns % GM June 2007
% That's all folks!
\end{document}